\documentclass{article}

\usepackage{natbib}
\usepackage{hyperref}
\usepackage{dirtytalk}
\usepackage{float}
\usepackage{graphicx}

\usepackage{amsmath}
\usepackage{amssymb}
\usepackage{bm}

\usepackage{color}
\usepackage{mathtools}

\definecolor{tomato}{RGB}{255, 99, 71}
\definecolor{lightblue}{RGB}{32, 114, 247}

\makeatletter
\def\blfootnote{\xdef\@thefnmark{}\@footnotetext}
\makeatother

\title{Moderated Network Models}
\date{}
\author{Jonas Haslbeck 	\qquad Denny Borsboom \qquad Lourens Waldorp \\ \\ \large \emph{Psychological Methods Group, University of Amsterdam}}

\begin{document}
	
\maketitle

\vspace{.8cm}

\begin{abstract}
	Pairwise network models such as the Gaussian Graphical Model (GGM) are a powerful and intuitive way to analyze dependencies in multivariate data. A key assumption of the GGM is that each pairwise interaction is independent of the values of all other variables. However, in psychological research this is often implausible. In this paper, we extend the GGM by allowing each pairwise interaction between two variables to be moderated by (a subset of) all other variables in the model, and thereby introduce a \emph{Moderated Network Model} (MNM). We show how to construct MNMs and propose an $\ell_1$-regularized nodewise regression approach to estimate it. We provide performance results in a simulation study and show that MNMs outperform the split-sample based methods Network Comparison Test (NCT) and Fused Graphical Lasso (FGL) in detecting moderation effects. Finally, we provide a fully reproducible tutorial on how to estimate MNMs with the R-package \emph{mgm} and discuss possible issues with model misspecification.
\end{abstract}	
	
	\blfootnote{This article has been accepted for publication in Multivariate Behavioral Research,
		published by Taylor \& Francis.}
	
	
\section{Introduction}

Network (or graphical) models are a powerful and intuitive tool to analyze dependencies in multivariate data and their popularity has recently surged in psychological research (e.g., personality psychology \citep{costantini2015state}, social psychology \citep{dalege2016toward}, and clinical psychology \citep{borsboom2017network, haslbeck2017predictable, eidhof2017replicability, kendler2018centrality}). The network models used in this literature model interactions between \emph{pairs} of variables, for instance \emph{mood} and \emph{physical activity}. Examples for such pairwise network models are the Gaussian Graphical Model (GGM) implied by the multivariate Gaussian distribution \citep{lauritzen1996graphical}, the Ising model for binary-valued data \citep{wainwright2008graphical} and the Mixed Graphical Model (MGM) for data from mixed distributions \citep{chen2014selection, yang2014mixed}.

A key assumption of these pairwise network models is that there are no moderation effects, which means that the interaction between any pair of variables is independent of the values of \emph{all other} variables in the network. As an example, let's say we have the variable \emph{fatigued} in the model. Then this assumption says that the (possibly positive) interaction between \emph{mood} and \emph{physical activity} does not depend on how fatigued an individual is. This is an empirical question and so seems worth checking. Because psychology studies highly contextualized and multicausal phenomena, the occurrence of such moderation effects is often plausible. Psychological researchers have known this for a long time, which is reflected by the widespread use of moderation in the analysis of psychological data: either explicitly as moderation analysis in the regression framework \citep[e.g.][]{fairchild2010evaluating, mackinnon2008and} or as interaction terms, for instance, in a 2-way Analysis of Variance \citep[e.g.][]{tabachnick2007using}.

Moderation is important in network models for the same reason it is important in these well-known analyses: missing a moderator means getting an average model (over the values of the missed moderator) that is potentially inappropriate for individuals with \emph{any} value on the moderator variable. This averaging can lead to detecting no effect at all, even if there is a strong moderation effect (see Section \ref{sec_methods_simplemoderation}). In addition, differences in the distributions of the missed moderator variable across studies offer one explanation for contradicting evidence across studies \citep{open2015estimating, open2012open}: for example, study A might have studied mostly well-rested individuals and found a positive relationship between \emph{mood} and \emph{physical activity}, while study B might have studied mostly fatigued individuals and found a negative relationship. Moderators are also important in clinical practice: for example, \cite{borsboom2017network} suggested that one can intervene on a symptom network of a mental disorder both on the node- and the network level. Since moderators determine the relationships between variables, they are the natural tool to find possible interventions on the network. More generally, moderators are useful in applications if a medication/treatment only works for a certain group of patients (defined by the moderator variable) and does not work (or even has adverse effects) for other groups of patients. Thus, studying moderator variables is a key requirement for moving towards personalized medicine and therapy \citep[e.g.][]{hamburg2010path}.

One way to check for moderation effects in network models is to split the data set in two parts along the moderator variable, estimate a network model on each of them, and compare them. A more sophisticated version of this procedure is the Network Comparison Test (NCT) \citep{van2016comparing} which performs a permutation test on differences across data sets for each edge-parameter; another procedure is the Fused Graphical Lasso (FGL), which jointly estimates two Gaussian Graphical Models (GGMs) on two data sets, with an additional penalty on differences between the two GGMs \citep{danaher2014joint}. These data-split approaches have two major draw-backs: first, the type of moderation effect they approximate is a step function, with the step placed at the value of the moderator variable at which the data set was split. Such a step function is implausible in many situations. In the above example, this would mean that the relationship between \emph{mood} and \emph{physical activity} remains (possibly positively) constant while increasing the value of \emph{fatigue}. And at some specific value of \emph{fatigue}, the relationship "jumps" to another (possibly negative) constant. Second, splitting the data set in half means losing information, because now both network models have to be estimated on half the data compared to the original data set. This leads to greatly reduced sensitivity to detect both pairwise interactions and moderation effects.

We propose a more direct approach to detecting moderation. Specifically, we extend pairwise network models by a specified set of moderator variables to obtain a \emph{Moderated Network Model} (MNM). This circumvents the above mentioned problems of the split-data approaches by fitting a linear moderation effect and making full use of the data. We do this by using the standard moderation definition from multiple linear regression and extend the multivariate Gaussian distribution with moderation effects. In a similar way one could also extend Ising models and more generally, MGMs, with moderation effects. Here, we take the first step by extending the popular GGM with moderation effects. Specifically, we make the following contributions:

\begin{enumerate}
	\item We introduce moderation for network models by extending pairwise network models with moderation effects similar to moderation effects in the linear regression framework
	\item We suggest a new visualization of moderated network models, based on factor graphs
	\item In a simulation study, we investigate the performance of the moderated network model in estimating moderation effects and compare it to the performance of the sampled-split based methods NCT and FGL
	\item A fully reproducible tutorial demonstrates how to fit and visualize Moderated Network Models using the R-package \emph{mgm}
\end{enumerate}

In Section \ref{sec_MNM}, we briefly review moderation in linear regression (\ref{sec_methods_simplemoderation}) and then show how to construct a Moderated Network Model (\ref{sec_construction}). In the last two subsections of Section \ref{sec_MNM} we show how to visualize (\ref{sec_intro_viz_modNW}) and estimate MNMs (\ref{methods_nodewisereg}) using an $\ell_1$-regularized nodewise regression approach. In Section \ref{sec_simulation} we report the performance of our estimation approach in estimating MNMs, and compare its performance in recovering moderation effects to the split-sample methods NCT and FGL. Finally, in Section \ref{sec_dataexamples}, we provide a fully reproducible tutorial (\ref{sec_tutorial}) on how to estimate MNMs with the R-package \emph{mgm} and discuss possible issues with model misspecification (\ref{sec_misspec}).

\section{Moderated Network Models}\label{sec_MNM}

In this section, we first review basic concepts of moderation in multiple regression, which are useful for introducing MNMs (Section \ref{sec_methods_simplemoderation}). Using these concepts, in Section  \ref{sec_construction} we construct the MNM by extending the multivariate Gaussian with 3-way interactions. In Section \ref{sec_intro_viz_modNW}, we show how to visualize MNMs using factor graphs and in Section \ref{methods_nodewisereg}, we present an $\ell_1$-regularized nodewise regression approach to estimate MNMs.

\subsection{Moderation in Linear Regression}\label{sec_methods_simplemoderation}

Here we review basic concepts of moderation in multiple regression, which we use to construct MNMs. Readers who are familiar with these concepts can skip directly to Section \ref{sec_construction}. 

\subsubsection{Moderation and Interactions in Linear Regression}

By moderation we mean that the effect of the predictor $B$ on response variable $A$ is a \emph{linear} function of a third variable $C$. The simplest possible example to introduce moderation is a linear regression model in which $A$ is a function of $B$ and $C$

\begin{equation}\label{eq_multiple_regression}
A = \beta_{B} B + \beta_{C} C + \varepsilon, 
\end{equation}

\noindent
where $\beta_{B}$ is the effect of $B$ on $A$, $\beta_{C}$ is the effect of $C$ on $A$, and  $\varepsilon$ has a Gaussian distribution with mean $\mu = 0$ and variance $\sigma^2$ \citep[e.g.][]{aiken1991multiple}. In this model, both effects $\beta_{B}, \beta_{C}$ are constants and therefore not a function of any variable. This changes when adding the product interaction term $BC$ with parameter $\omega_{BC}$ as a predictor to the model

\begin{align} 
A &= \beta_{B} B + \beta_{C} C + \omega_{BC} BC + \varepsilon  \label{eq_multiple_regression_interaction_1} \\ 
&= ( \beta_{B} + \omega_{BC} C) B + \beta_{C} C + \varepsilon \label{eq_multiple_regression_interaction_2} \\ 
&= \beta_{B} B + ( \beta_{C} + \omega_{BC} B) C + \varepsilon .  \label{eq_multiple_regression_interaction_3}
\end{align}

\noindent
Rewriting the model with interaction the term in (\ref{eq_multiple_regression_interaction_1})  into (\ref{eq_multiple_regression_interaction_2}) shows that the effect of $B$ on $A$ is now equal to

\begin{equation} \label{eq_mod_main_effect}
(\beta_{B} + \omega_{BC} C),
\end{equation}

\noindent and therefore a linear function of $C$, with constant term $\beta_{B}$ and slope $\omega_{BC}$. If the effect of $B$ on $A$ depends linearly on $C$, we say that this effect is linearly moderated by $C$. Because we can rewrite (\ref{eq_multiple_regression_interaction_1}) also into (\ref{eq_multiple_regression_interaction_3}), $\omega_{BC}$ can also be interpreted as the moderation effect of $C$ on the effect of $B$ on $A$. The above rewriting shows that the interaction effect and moderation effects are different interpretations of the same parameter $\omega_{BC}$. Throughout the paper we adopt the moderation perspective, because it is more intuitive and the parameter is easier to interpret.

To develop some intuition for moderation, let's consider an example in which $\beta_B = 0.2$ and $\omega_{BC} = 0.1$. Then, if $C=0$ the effect of $B$ on $A$ is equal to $0.2 + 0.1 \cdot 0 = 0.2$. And if $C=1$, it is equal to $0.2 + 0.1 \cdot 1 = 0.3$. In this example the effect of $B$ on $A$ is equal to the constant  $\beta_B$ plus $C$ times $\omega_{BC}$. Figure \ref{fig_moderation_simple} shows four possible cases for effects of $B$ on $A$:

\begin{figure}[H]
	\centering
	\includegraphics[width=.9\textwidth]{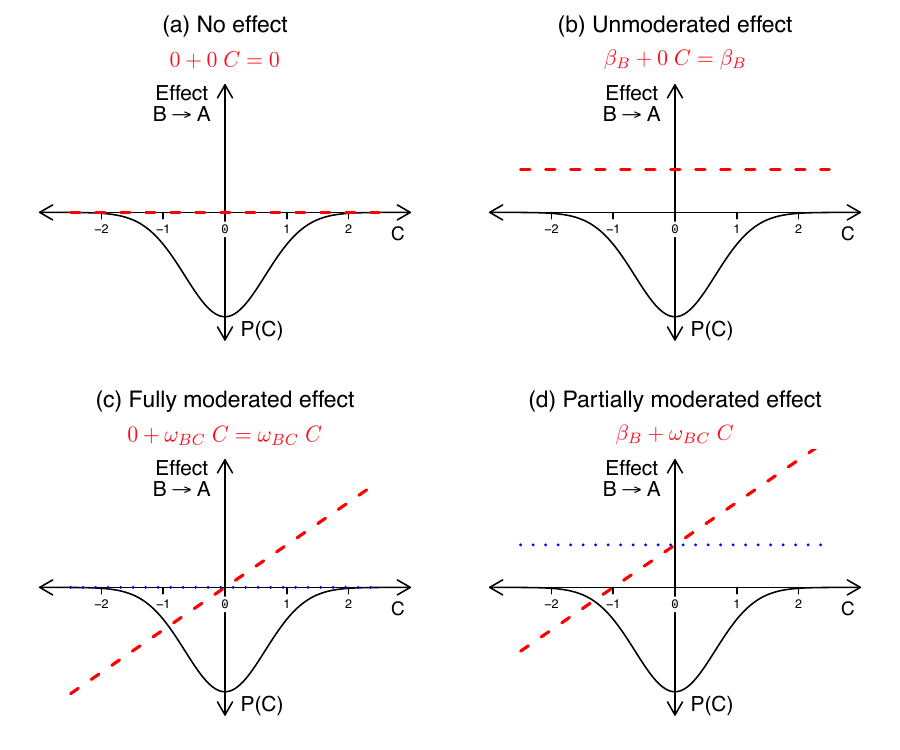}
	\caption{The linear function in equation (\ref{eq_mod_main_effect}) determining the main effect of $B$ on $A$ for the four possible zero/nonzero combinations of the parameters $\beta_{B}$ and $\beta_{BC}$, displayed both as equation and visually (dashed red line). The dotted blue line in panels (c) and (d) indicate the effect one would obtain when fitting a regression model without moderation/interaction term.}\label{fig_moderation_simple}
\end{figure}

The x-axis shows the values of the moderator $C$. The y-axis shows both the effect of $B$ on $A$ as a function of $C$ and the density of $C$. In panel (a) of Figure \ref{fig_moderation_simple} there is no effect of $B$ on $A$; in panel (b) there is a constant effect from $B$ on $A$ that is independent of $C$; in panel (c) there is an effect of $B$ on $A$ that is fully determined (moderated) by $C$; and in panel (d) the effect of $B$ on $A$ is equal to a constant plus a dependency on $C$. The dotted blue lines in panel (c) and (d) of Figure \ref{fig_moderation_simple} indicate the parameter one would obtain for the effect of $B$ on $A$ with a simple regression model \emph{without} moderation/interaction term. We make two observations: first, the constant parameters (dotted blue line) are a poor description of the true moderated parameters (dashed red line). Second, (c) shows that if $\beta_{B} = 0$ and $|\omega_{BC}|>0$, one would entirely miss the presence of an effect when estimating a regression model without moderation.

In the procedure to estimate MNM described in Section \ref{methods_nodewisereg} we mean-center all variables before estimation. This is required to ensure that all parameters in the MNM have a meaningful interpretation. We discuss this issue in Appendix \ref{sec_linreg_centering}, or refer the reader to \cite{afshartous2011key}.

\subsubsection{Regression vs. Network Semantics}\label{sec_semantics}

In the following section we wed the world of regression with the world of networks. Unfortunately, depending on whether one adopts the regression (conditional distribution) or network (joint distribution) perspective, the same parameter is referred to differently. To avoid confusion, we make these differences explicit: In Section \ref{sec_methods_simplemoderation}, we discussed moderation from the regression (conditional distribution) perspective. In regression, $\beta_{B}$ is typically referred to as main effect, or conditional main effect in the presence of moderation, and $\omega_{BC}$ is referred to as the moderation/interaction effect. From the perspective of the network model, $\beta_{B}$ is referred to as a pairwise interaction (because it is associated with the product $AB$ in the joint distribution), and $\omega_{BC}$ is referred to as a 3-way interaction (because it is associated with the product $ABC$ in the joint distribution) or a moderation effect (because it moderates the pairwise interaction $AB$). The two different perspectives will become apparent in the following section, where we show both the joint distribution of the MNM and the conditional distributions. In the remainder of the paper we adopt the network perspective, except when otherwise stated.

\subsection{Gaussian Distribution and Gaussian Graphical Model}

A graphical (or network model) is a statistical model for which an undirected graph/network encodes the conditional dependence structure between random variables \citep{lauritzen1996graphical}. A popular network model for continuous data is the multivariate Gaussian distribution. We introduce this distribution here, because we will use it as a basis for constructing Moderated Network Models for continuous data in the following section:

\begin{equation}\label{eq_standardpGaussian}
	P(X = x) = \frac{1}{\sqrt{(2 \pi)^p|\Sigma|}} 
	\exp \left \{
	- \frac{1}{2} 
	(x - \mu)^\top
	\Sigma^{-1}
	(x - \mu)  
	\right
	\}
	,
\end{equation}

\noindent
where $x$ is $p$-dimensional vector of random variables, $\mu$ is a $p$-dimensional vector of means, $\Sigma$ is a $p \times p$ variance-covariance matrix, and $|\Sigma|$ denotes the determinant of $\Sigma$.

In the case of the multivariate Gaussian distribution, it is easy to obtain the graph/network that encodes the conditional dependence structure between random variables: if an entry in the inverse variance-covariance matrix $\Sigma^{-1}$ is nonzero, the two corresponding variables are conditionally dependent (present edge in the network); if the entry is zero, the two corresponding variables are conditionally independent (no edge in the network). The resulting conditional (in-)dependence network is also called Gaussian Graphical Model (GGM). For an accessible introduction to GGMs, their relation to regression and Structural Equation Modeling (SEM), and how to estimate them, we refer the reader to \cite{epskamp2018gaussian}.

\subsection{Construction of Moderated Network Model}\label{sec_construction}

The central goal of this paper is to construct a joint distribution over $p$ continuous variables which allows that each pairwise interaction between variables $X_i$ and $X_j$ is a linear function of all other variables. Another way of saying this is that each pairwise interaction between $X_i$ and $X_j$ is linearly moderated by all other variables. Here we construct such a joint distribution by adding moderation effects to the multivariate Gaussian distribution, which models (unmoderated) linear pairwise interactions between variables.

The density of the multivariate Gaussian distribution (\ref{eq_standardpGaussian}) shown in the previous section can be rewritten in its exponential family form as

\begin{equation}\label{eq_multivariateGaussian}
	P(X) = 
	\exp
	\left
	\{
	\sum_{i}^p \alpha_i \frac{X_i}{\sigma_i} + 
	\sum_{\substack{i,j \in V\\ i \neq j}} \beta_{i,j} \frac{X_i}{\sigma_i} \frac{X_j}{\sigma_j} + 
	\sum_{i}^p \frac{X_i^2}{\sigma_i^2} - \Phi(\alpha, \beta) 
	\right
	\}
	,
\end{equation}

\noindent
where $X \in \mathbb{R}^p$, $V = \{1, 2, \dots, p\}$ is the index set for the $p$ variables, $\alpha$ is a $p$-vector of intercepts, $\beta$ is a $p \times p$ matrix of ${p}\choose{2}$ partial correlations\footnote{${{n}\choose{k}} = \frac{n!}{k!(n-k)!}$, the $k^{\text{th}}$ binomial coefficient of the polynomial expansion of $(1+x)^n$.}, $\sigma_i^2$ is the variance of $X_i$, and $\Phi(\alpha, \beta)$ is the log normalizing constant which ensures that the probability distribution integrates to 1. 

To see how the common form of the Gaussian density in equation (\ref{eq_standardpGaussian}) can be written as equation (\ref{eq_multivariateGaussian}) above, momentarily assume that all means are equal to zero $\mu = 0$. Then the term in the exponential simplifies to $- \frac{1}{2} x^\top \Sigma^{-1} x$. This inner product is the same as the sum 	$\sum_{\substack{i,j \in V\\ i \neq j}} \beta_{i,j} \frac{X_i}{\sigma_i} \frac{X_j}{\sigma_j}$ in (\ref{eq_multivariateGaussian}), except that we summed over each $(i, j)$ combination twice, which is why we multiply with $\frac{1}{2}$. With means unequal zero, one can expand the expression in the exponential and gets a similar expression for the interactions, plus an expression for the intercepts $\alpha$, which are a function of the means and the interaction parameters.

Now, to extend the Gaussian distribution in (\ref{eq_multivariateGaussian}) with all possible moderation effects, we add all 3-way interactions to the model:

\begin{align} \label{eq_MNM_joint}
P(X) =
\exp&
\left \{
\sum_{i}^p \alpha_i \frac{X_i}{\sigma_i} + 
\sum_{\substack{i,j \in V\\ i \neq j}} \beta_{i,j}  \frac{X_i}{\sigma_i} \frac{X_j}{\sigma_j} + \right. \nonumber \\ & \left.
\sum_{\substack{i,j,q \in V\\ i \neq j \neq q}} \omega_{i, j, q} \frac{X_i}{\sigma_i} \frac{X_j}{\sigma_j} \frac{X_q}{\sigma_q} + 
\sum_{i}^p \frac{X_i^2}{\sigma_i^2} - \Phi(\alpha, \beta, \omega) 
\right \} 
,
\end{align}

\noindent
where $\omega$ is a $p \times p \times p$ array of ${p}\choose{3}$ 3-way interactions. 

How many parameters are introduced by adding 3-way interactions? For $p=10$ variables, adding all 3-way interactions means that the number of interaction parameters increases from ${{10}\choose{2}} = 45$ to ${{10}\choose{2}} + {{10}\choose{3}} = 45 + 120 = 165$. Instead of adding all moderation effects (3-way interactions) one can also add single moderation effects, or all moderation effects of a subset of variables. For instance, adding all moderation effects of $M \in \{1, 2, \dots, p\}$ moderators would result in $\sum_{m=1}^M \sum_{i=1}^{\min\{0, p-1-m\}} i$ additional moderation parameters.

The distribution of $X_s$ conditioned on all remaining variables $X_{\setminus s}$ is given by

\begin{align}\label{eq_conditional_MNM}
P(X_s | X_{\setminus s}) =
\exp&
\left \{
\alpha_s \frac{X_s}{\sigma_s} + 
\sum_{\substack{i \in V\\ i \neq s}} \beta_{i,s}  \frac{x_i}{\sigma_i} \frac{X_s}{\sigma_s} + \right. \nonumber \\ & \left.
\sum_{\substack{i,j \in V\\ i \neq j \neq s}} \omega_{i, j, s} \frac{x_i}{\sigma_i} \frac{x_j}{\sigma_j} \frac{X_s}{\sigma_s} + 
\frac{X_s^2}{\sigma_s^2} - \Phi^*(\alpha, \beta, \omega) 
\right \} 
,
\end{align}

\noindent
where $\Phi^*(\alpha, \beta, \omega)$ is the log-normalizing constant with respect to the conditional distribution $P(X_s | X_{\setminus s})$, and we use lower case letters $x_i$ to indicate fixed values opposed to random variables $X_i$. We now show that (\ref{eq_conditional_MNM}) is a conditional Gaussian distribution. To make the presentation more clear, we set $\sigma_s = 1$ without loss of generality. If we let

\begin{equation}\label{eq_conditional_mean}
	\mu_s = \alpha_s + \sum_{\substack{i \in V\\ i \neq s}} \beta_{i,s}  x_i + 
	\sum_{\substack{i,j \in V\\ i \neq j \neq s}} \omega_{i, j, s} x_i x_j
\end{equation}

\noindent
and

$$
\Phi^*(\alpha, \beta, \omega) = \frac{\mu_s}{2} - \log (\frac{1}{\sqrt{2 \pi }})
$$

\noindent
we can rewrite (\ref{eq_conditional_MNM}) into 

\begin{align}\label{eq_conditional_MNM_rewritten}
P(X_s | X_{\setminus s}) =&
\frac{1}{\sqrt{2 \pi }}
\exp
\left \{
\mu_s X_s - \frac{X_s}{2} - \frac{\mu_s}{2}
\right \} \\ \nonumber
=&
\frac{1}{\sqrt{2 \pi }}
\exp
\left \{
\frac{( X_s - \mu_s)^2}{2}
\right \}
,
\end{align}

\noindent
which is the well-known form of the conditional Gaussian distribution. In Appendix \ref{app_HOI_and_norm}, we provide more intuition for the MNM by deriving the joint distribution in (\ref{eq_MNM_joint}) from the conditionals as in (\ref{eq_conditional_MNM_rewritten}) for the case of $p=3$ variables.

Since each $P(X_s | X_{\setminus s})$ is only parameterized by $\mu_s$ and regression estimates $\mathbb{E}[X_s | X_{\setminus s}] = \mu_s$, the above derivation shows that a Moderated Network Model can be estimated with a series of $p$ regressions that include the appropriate moderation (3-way interaction) effects. Specifically, the equation for the mean $\mu_s$ of the conditional distribution $X_s$ in (\ref{eq_conditional_mean}) has the same form as the moderated linear regression in equation (\ref{eq_multiple_regression_interaction_1}) in Section \ref{sec_methods_simplemoderation}, except that it includes more terms. In Section \ref{methods_nodewisereg}, we show how to estimate the $p$ conditional Gaussians using $\ell_1$-regularized regression and how to combine the resulting estimates to the MNM joint distribution.

The mean (\ref{eq_conditional_mean}) of the conditional distribution $P(X_s | X_{\setminus s})$ compared to the joint distribution $P(X)$ in (\ref{eq_MNM_joint}) explains the different terminology for interaction effects, depending on whether one adopts the regression- or graph perspective (see Section \ref{sec_semantics}). For example, in the joint distribution, the second term indicates pairwise interactions because two variables are multiplied. In the mean of the conditional, which is estimated in a linear regression model, the second term only contains a single variable and is therefore referred to as a main effect.

Above we discussed that one could include all 3-way interactions or only a subset of 3-way interactions in the model. However, we always include all pairwise interactions. While all pairwise interactions are included in the model, this does not mean that the parameter associated with a pairwise interaction has to be nonzero if the 3-way interaction moderating that pairwise interaction is nonzero. In other words, in the joint MNM (\ref{eq_MNM_joint}) the presence of a 3-way interaction does not imply the presence of a pairwise interaction. This is in contrast to log-linear models for categorical data, in which a $k$-order interaction always implies a $(k-1)$-order interaction \citep[e.g.][]{agresti2011categorical}.

The number of parameters of the MNM is much larger than in pairwise network models, especially when $p$ is large. If the proportion of nonzero 3-way interaction parameters in the true model would be the same as the proportion of nonzero pairwise interaction in the true model, we needed a lot more observations $n$ to estimate the MNM with similar accuracy as the pairwise model. However, it is highly implausible that \emph{every} variable in the model moderates \emph{every} pairwise interaction. Instead, we would expect that \emph{some} variables moderate \emph{some} pairwise interactions. Under the assumption that a large fraction of moderation effects are equal to zero in the true model, it is possible to estimate the model accurately with a reasonable sample size $n$. In Section \ref{methods_nodewisereg}, we present an $\ell_1$-regularized regression procedure to estimate MNMs, which uses this assumption. In Section \ref{sec_simulation}, we explicitly show in a simulation study how much data is needed to recover a MNM in realistic situations.


We showed that the MNM joint distribution in (\ref{eq_MNM_joint}) can be factorized into $p$ conditional Gaussian distributions. However, the MNM joint distribution is not a multivariate Gaussian distribution, because we added 3-way interactions (moderation effects). For the multivariate Gaussian distribution, all parameters have to be finite and the covariance matrix has to be positive-definite to ensure that the distribution is normalizable. For the MNM proposed here, the constraints to ensure normalizability are unknown. For the class of MGMs, which generalize MNMs, \citep{yang2014mixed} proposes that for normalizability it is sufficient that the sum of unnormalized terms in the exponential in (\ref{eq_MNM_joint}) are smaller than zero, which will ensure that the unnormalized mass in (\ref{eq_MNM_joint}) converges to zero. Thus, in order to ensure normalizability, one needs to constrain the parameter space such that this inequality is satisfied. The required constraint is most likely a constraint on the 3-way interactions, and is a function of all other parameters and the variances of the conditional Gaussian distributions. However, working out these constraints is beyond the scope of this paper, in which we focus on introducing the idea of moderation in network models to an applied audience. In the present paper, we therefore estimate MNMs with an unconstrained nodewise algorithm. This means that we do not know whether the estimated parameters lead to a normalizable joint distribution. One consequence of having no guarantee that the joint distribution is normalizable is that one cannot apply any global goodness of fit analyses, for example, to select between models with different sets of included moderators. While this is a major limitation of the here proposed MNM, one can perform model selection with out of sample prediction error, which does not require a proper joint distribution.


\subsection{Visualizing Moderated Network Models}\label{sec_intro_viz_modNW}

Pairwise network models are typically visualized in a network consisting of nodes representing variables and undirected edges representing pairwise interactions. In MNMs, we have additional moderation parameters and therefore need to find a new visualization that allows to include those without giving up clarity. We solve this problem with a factor graph visualization, in which each interaction (pairwise or higher-order) is represented by a \textit{factor node} \citep[see e.g.][]{koller2009probabilistic}. We first show how to represent a pairwise network model as a factor graph and then demonstrate how to include moderation parameters in the factor graph visualization.

\begin{figure}[H]
	\centering
	\includegraphics[width=1\textwidth]{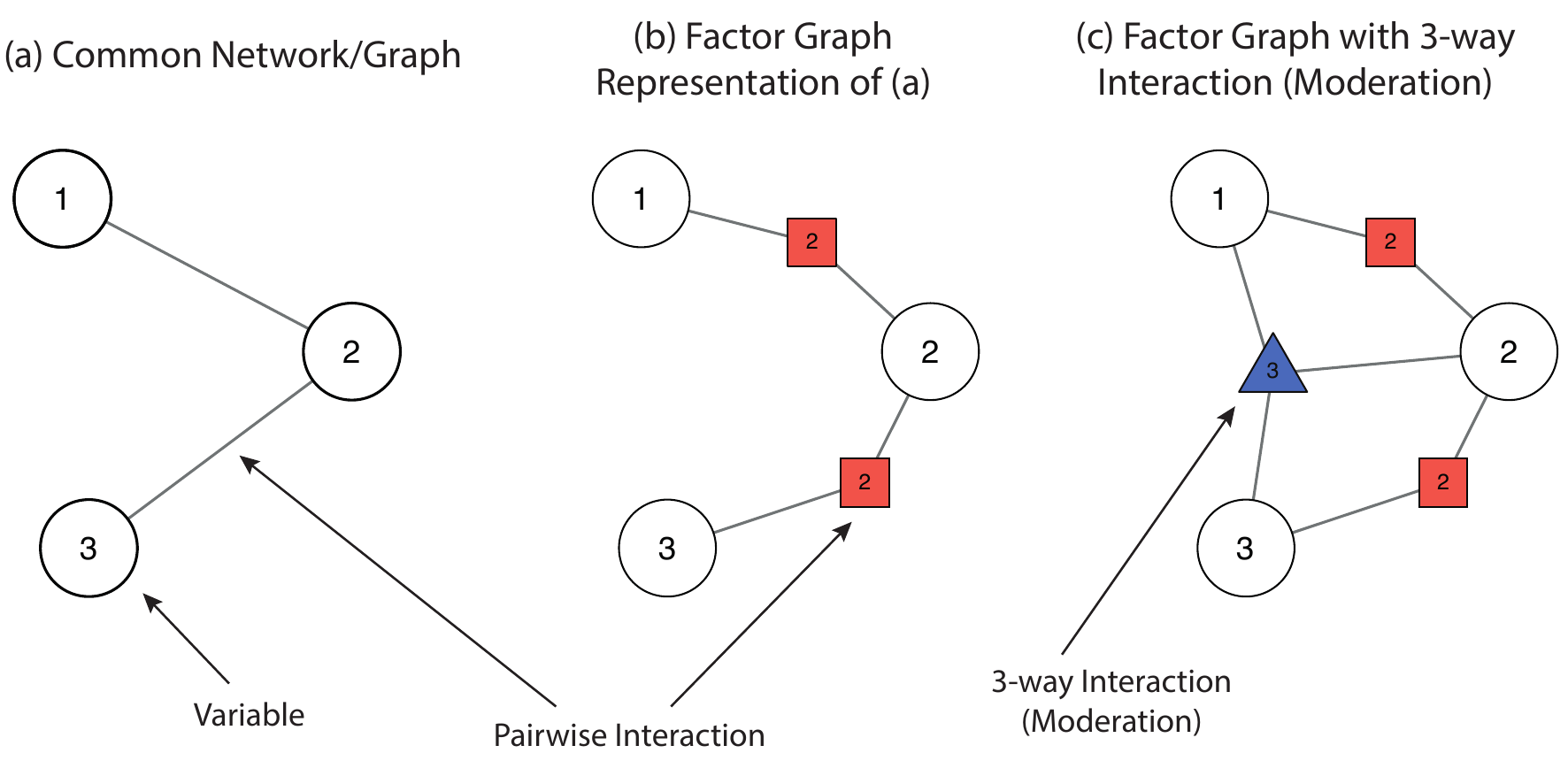}
	\caption{(a) The typical network-visualization of a network model with pairwise interactions between variables 1-2 and 2-3; (b) the network model in (a) visualized as a factor graph. Edges are now represented by factor nodes (red square nodes) of order 2; edges now indicate which variables are connected to which factor (c) the factor graph visualization of the previous network model with an additional 3-way interaction (blue triangle node).}\label{fig_intro_factorgraph}
\end{figure}

Figure \ref{fig_intro_factorgraph} (a) shows the typical network-visualization of a pairwise network model with edges 1-2 and 2-3 indicating pairwise interactions between those two pairs of variables. Panel (b) shows the visualization of the same network model as a factor graph: now each pairwise interaction is represented by a square red factor node which connects to the nodes that are involved in the respective pairwise interaction. In the present example the network model has two pairwise interactions, each of which is now represented by a factor nodes. The label 2 indicates that the interaction is pairwise. 

The factor graph visualization allows us to also include interactions involving three variables (3-way interactions): in panel (c) we add the 3-way interaction 1-2-3 to the pairwise model and visualize it in a factor graph. Again we visualize the two pairwise interactions 1-2 and 2-3 as separate factor nodes. Similarly, we visualize the 3-way interaction as a factor node that connects to the three variables involved in the 3-way interaction. Again, the label 3 indicates that the interaction is a 3-way interaction. We can interpret this 3-way interaction in terms of moderation in three different ways: (i) the moderation effect of 3 on the pairwise interaction 1-2, (ii) the moderation effect of 1 on the pairwise interaction 2-3, and (iii) the moderation effect of 2 on the pairwise interaction 1-3. In Section \ref{sec_methods_simplemoderation} we called (i) and (ii) \emph{partially} moderated because there is a pairwise interaction independent of the value of the moderator variable. And we called (iii) \emph{fully} moderated, because the pairwise interaction is fully determined by the value of the moderator variable.

In Figure \ref{fig_intro_factorgraph} we did not display the value of the two pairwise and the 3-way interaction. In a factor graph this information can either be displayed on the factor nodes or on the edges connecting them to variables. Since researchers are already familiar with displaying the weight of parameters as the width of edges, we chose the latter option. We show such weighted factor graphs in empirical data examples in Section \ref{sec_dataexamples}.

\subsection{Estimation via $\ell_1$-regularized Nodewise Regression}\label{methods_nodewisereg}

In this section we show how to estimate the $p$ conditional distributions of the MNM with $\ell_1$-regularized nodewise regression and how to combine the estimates to the joint MNM. Our approach is similar to the nodewise regression approach for estimating the multivariate Gaussian distribution \citep{meinshausen2006high}, except that we estimate the conditionals of the MNM instead of the joint Gaussian distribution.

\subsubsection{Estimate Nodewise Regressions}

To estimate the $p$ regression models, we minimize the squared loss plus the $\ell_1$-norm of the parameter vectors $\beta_{s, \cdot}$ and $\omega_{s, \cdot, \cdot}$ for each variable $s \in V$

\begin{equation}\label{eq_nw_squared}
\arg_{\beta_{s, \cdot}, \omega_{s, \cdot, \cdot}} \min \left \{  
\sum_{z=1}^n \ (X_{z, s} - \hat \mu_{z, s}  )^2
+ \lambda_s ( || \beta_{s, \cdot} ||_1 + ||\omega_{s, \cdot, \cdot} ||_1)
\right \} 
,
\end{equation}

\noindent
where $X_{z, s}$ is the value of variable $s$ in row $z$ of the data matrix, $\hat \mu_{z, s}$ is the predicted mean for row $z$ (see equation \ref{eq_conditional_mean}), $\beta_{s, \cdot}$ and $\omega_{s, \cdot, \cdot}$ are vectors containing parameters associated with pairwise and 3-way interactions, respectively,   $|| \beta_s ||_1 + || \omega_s ||_1 = \sum_{\substack{i \in V \\ i\neq s}} |\beta_{s,i}| + \sum_{\substack{i,j \in V \\ i \neq j \neq s}} |\beta_{s, i,j}|$ is the sum of the $\ell_1$-norms of both parameter vectors, and $\lambda_s$ is a tuning parameter that weights the $\ell_1$-norm relative to the squared loss. Note that $\beta_{s, \cdot}$ contains only the pairwise interactions involving variable $s$ and has therefore less elements than $\beta$ in the joint distribution. Similarly, $\omega_{s, \cdot, \cdot}$ contains only the 3-way interactions involving variable $s$ and has therefore less elements than $\omega$ in the joint distribution.

Prior to estimation we mean-center each variable and divide each variable by its standard deviation. This ensures that the penalization of a given parameter does not depend on the standard deviation of its associated (product of) variables and simplifies the model because all intercepts are equal to zero. In addition, recall that mean-centering of variables is necessary to obtain interpretable parameter estimates (see Section \ref{sec_linreg_centering}).

We chose $\ell_1$-regularized (LASSO) regression for three reasons: (1) the number of parameters can be large if $p$ is large and many moderation effects are included in the model (see Section \ref{sec_construction}), which leads to high variance on the parameter estimates (overfitting). The $\ell_1$-regularization shrinks parameter estimates towards zero and thereby mitigates this problem; (2) The $\ell_1$-penalty sets small parameter estimates to zero, which simplifies the interpretation of the model, especially when interpreted from a network/graph perspective; and (3) $\ell_1$-penalization ensures that the model remains identifiable when the number of parameters is larger than the number of observations. When estimating an MNM with $p=20$ variables and includes all moderation effects, each nodewise regression has $p-1 + {{p-1}\choose{2}} = 190$ parameters, which means unregularized methods require $n \geq 190$ observations. The $\ell_1$-penalization allows to estimate such a model also with $n < 190$.

The main assumption underlying $\ell_1$-regularized regression is that most of the parameters in the true model are equal to zero (also called \emph{sparsity} assumption). This seems a reasonable assumption for the MNM in psychological data, since we would not expect that every variable moderates every pairwise interaction. For an excellent discussion of $\ell_1$-regularized regression see \cite{hastie2015statistical}.

In each of the $p$ regressions, one has to select a tuning parameter $\lambda_s$ which controls the strength of the penalization. If $\lambda_s = 0$, the loss function in (\ref{eq_nw_squared}) reduces to squared loss alone, which is the loss function of standard OLS regression. If $\lambda_s$ is huge, all parameters are set to zero. To select an optimal $\lambda_s$, one can use a cross-validation scheme or an information criterion. \cite{foygel2010extended} showed that the Extended Bayesian Information Criterion (EBIC), a modification of the BIC \cite{schwarz1978estimating} that puts an extra penalty on nonzero parameters, performs well in estimating sparse parameter vectors.

\subsubsection{Combine Estimates to Joint Moderated Network Model}

The above estimation procedure leads to two estimates for every pairwise interaction, and three estimates for every 3-way interaction (moderation effect). For example, we obtain an estimate for the pairwise interaction between $X_i$ and $X_j$ from the regression of $X_i$ on $X_j$, and another estimate from the regression of $X_j$ on $X_i$. Similarly, we obtain three estimates for any given moderation effect from three regressions: the nodewise estimation procedure returns three estimates for each moderation parameter: (1) $X_q$ moderating the predictor $X_s$ in the regression on $X_j$, (2) $X_s$ moderating the predictor $X_j$ in the regression on $X_q$, and (3) $X_q$ moderating the predictor $X_j$ in the regression on $X_s$. In order to arrive at a single estimate to specify the joint MNM, we either take the arithmetic mean across the two/three values (OR-rule), or take the arithmetic mean across the two/three values if all three values are nonzero and otherwise set the aggregated parameter to zero (AND-rule). The AND-rule is more conservative than the OR-rule. It is even more conservative for 3-way interactions, because now three parameter estimates have to be nonzero to set the aggregate parameter to nonzero. For a more elaborate description of the nodewise regression procedure see \cite{haslbeck2017mgm}.

In the following section, we investigate the performance of the $\ell_1$-regularized nodewise regression approach in estimating MNMs, and compare its performance in detecting moderation effects to the split sample methods NCT and FGL.

\section{Simulation Study}\label{sec_simulation}

The goal of this simulation is (a) to investigate the performance of $\ell_1$-regularized nodewise regression in estimating moderated network models and (b) compare its performance to detect moderation effects to the split-sample methods Network Comparison Test (NCT) and Fused Graphical Lasso (FGL). Note that since the NCT and FGL can only provide a piecewise constant approximation of the linear moderation effects in MNMs, we expect that they will perform worse than MNMs. However, because the methods differ in several additional characteristics, and to determine the exact performance differences, we map out the differences of NCT, FGL and MNMs. We first describe the data generation (\ref{sec_simulation_datageneration}) and the estimation procedures (\ref{sec_sim_est}). Finally, we report performance results (\ref{sec_sim_results}) and discuss them in Section \ref{sec_sim_discussion}.

\subsection{Data generation}\label{sec_simulation_datageneration}

We sample observations from 100 Moderated Network Models that are specified by the following procedure: we begin with  an empty graph with $p=12$ nodes and randomly add six edges. Of these six edges, two are unmoderated pairwise interactions (e.g. edge 2-4 in Figure \ref{fig_sim_generating}; or panel (b) in Figure \ref{fig_moderation_simple}), two are fully moderated pairwise interactions (e.g., 6-7 in Figure \ref{fig_sim_generating}; or panel (d) in Figure \ref{fig_moderation_simple}), and two are partially moderated pairwise interactions (e.g. edge 12-1 in Figure \ref{fig_sim_generating}; or panel (d) in Figure \ref{fig_moderation_simple}). We enforce that each node has at most 2 edges by resampling the graph until this constraint is met. We do this because sampling from highly connected nodes leads to many rejections in the rejection sampler, which makes sampling unfeasible. After obtaining the graph with six edges, we add an additional variable, which serves as the moderator\footnote{We do not allow the initial six edges to be connected to the moderator, because otherwise for some graph configurations (1) moderation effects can turn into quadratic effects and (2) unmoderated pairwise interactions can turn into moderated pairwise interactions.}. The final graph therefore has 13 nodes. We repeat this procedure for 100 iterations, yielding 100 data generating models in the simulation. Figure \ref{fig_sim_generating} shows the model resulting from this procedure in iteration 2:

\begin{figure}[H]
	\centering
	\includegraphics[width=.9\textwidth]{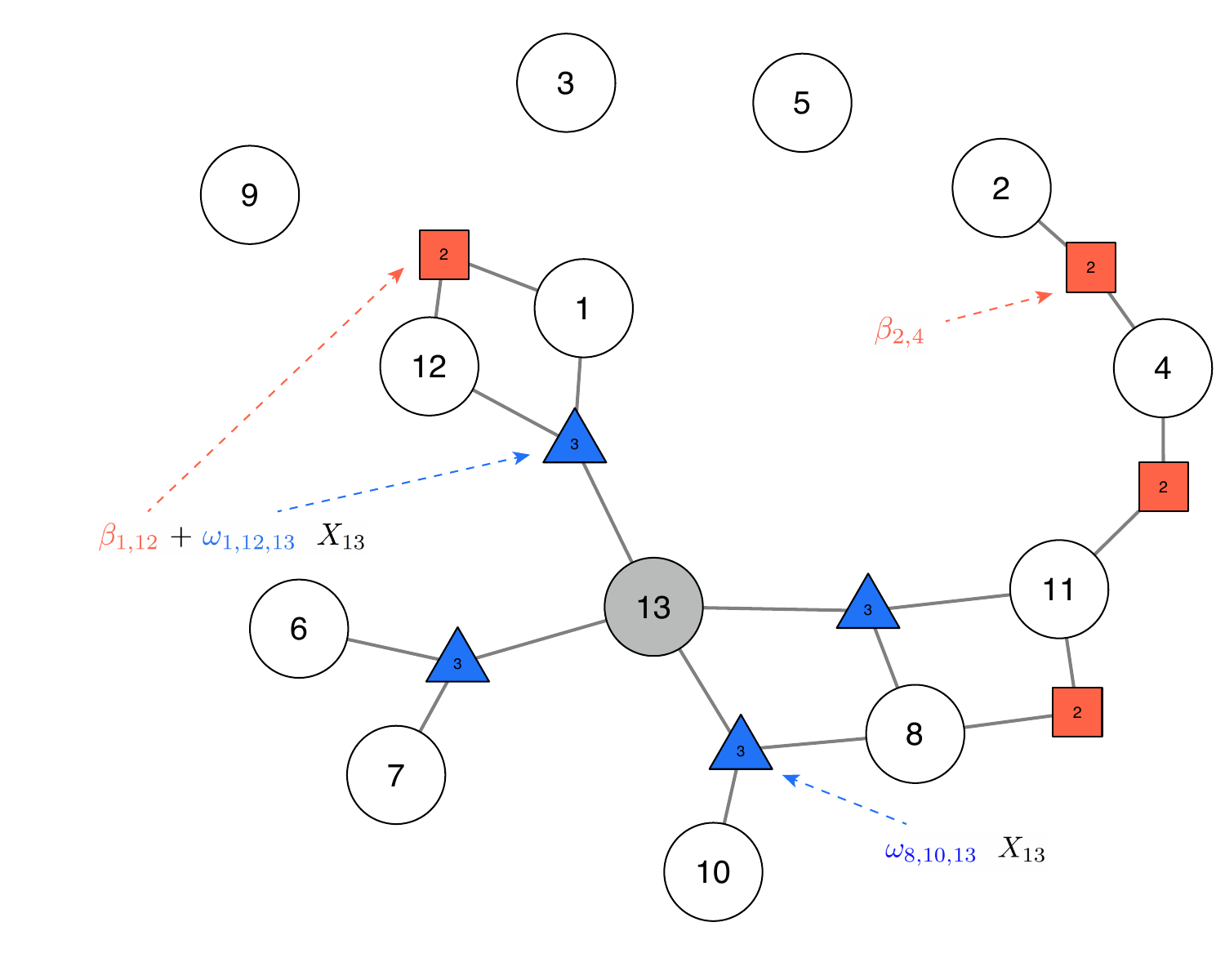}
	\caption{The factor graph used in iteration 2 of the simulation. Circle nodes indicate variables. Square nodes with label 2 indicate 2-way interactions. Triangle nodes with label 3 indicate moderation effects (3-way interactions). Each of the three types of pairwise interactions (unmoderated, partially moderated, fully moderated) appears twice. For one of each of them the formula for the total effect is shown. Node 13 is the moderator variable.}\label{fig_sim_generating}
\end{figure}

We visualize the moderated network model using a factor graph (see Section \ref{sec_intro_viz_modNW}). There is an unmoderated pairwise interaction between 2-4 and 4-11; a partially moderated pairwise interaction between 1-12 and 8-11; and a fully moderated pairwise interaction between 6-7 and 8-10. All moderated interactions are moderated by variable 13. The factor graph visualization highlights the equivalence between moderation effects and higher order interactions: a pairwise interaction between $A$ and $B$ that is partially moderated by $C$ means that $A$ and $B$ are also connected to $C$ in a 3-way interaction. Take the pairwise interaction 1-12 as an example: the total pairwise interaction is equal to the constant $\beta_{1,12}$ plus variable $X_{13}$ weighted by the moderation effect $\omega_{1,12, 13}$. The presence of a 3-way interaction alone can be seen as full moderation: for example 8-10 are only connected via the 3-way interaction with 13. This means the parameter for the pairwise interaction 8-10, is only a function of $X_{13}$ weighted by the moderation effect $\omega_{6,7,13}$. In an unmoderated pairwise interaction (for example 2-4) the two variables are not involved in the same 3-way interaction. In this case, the total parameter for the pairwise interaction is a constant.

We constructed the joint distribution by factoring $p$ conditional Gaussians (see Section\ref{sec_construction}). For each of the $p$ conditional Gaussians, we set the standard deviation to one and the intercept to zero. To be able to compare the performance in recovering pairwise and 3-way interactions, we set the value of all nonzero parameters $\beta_{i,j}$ and $\omega_{i,j,q}$ to $0.2$. The total interaction parameters $(\beta_{i,j} + \omega_{i,j,q} X_q)$ can be interpreted as (moderated) partial correlations. To sample cases from the joint distribution, we use a Gibbs sampler on the $p$ conditional Gaussians with means defined in equation (\ref{eq_conditional_mean}). As discussed in Section \ref{sec_construction},  the constraints of the parameter space under which the joint distribution is normalizable are unknown. We work around this problem by using a rejection sampler (for details see Appendix \ref{app_Sampling}). With this sampling procedure we obtain $n = 1808$ cases from each of the 100 MNMs.

To investigate performance as a function of the number of observations $n$ we create 12 variations on a log scale $n \in \{30, 46, \dots, 1148, 1808 \}$. This range of $n$-values was chosen because it allows us to show the performance transition from detecting no parameters at all to perfectly recovering the model as a function of $n$. We always use the first $30, 46, \dots$ observations, which means that in scenario $n = 46$ we take the samples of the scenario $n = 30$ and add the next 16. This approach minimizes differences in performance across $n$-variations due to sampling variation.

\subsection{Estimation}\label{sec_sim_est}

Here we describe the three different methods for detecting moderation effects that we compare in the simulation study: Moderated Network Models (\ref{sec_sim_est_MNM}), the Netwok Comparison Test (NCT) (\ref{sec_sim_est_NCT}), and the Fused Graphical Lasso (FGL) (\ref{sec_sim_est_FGL}).

\subsubsection{Moderated Network Models}\label{sec_sim_est_MNM}

To investigate the performance of MNMs in recovering moderation effects we estimate MNMs in three different versions: in version (1) we know the true moderator variable (variable 13) and specify only that variable as a moderator. If we do not know the true moderator, one can use two different strategies: in version (2) we estimate $p$ moderated network models, in each of which we specify a different variable as the moderator. After estimating the $p$ models, all estimates are combined such that if a parameter was estimated nonzero in at least one of the $p$ models it is considered to be present in the combined model; note that the sensitivity to discover pairwise interactions for (2) will be larger or equal compared to (1) because we combine 13 estimates for each pairwise interaction. The same is true for moderation effects, but we expect the difference to be smaller, because only 3 estimates are combined (in our model the moderation effect of $C$ on the interaction between $B,A$ is the same as the moderation effects of $B$ on the interaction between $A, C$ and of $A$ on the interaction between $C, B$).  The reason to include this version of the algorithm is to see how much the precision drops when using such an exploratory approach. The second strategy for the situation in which the true moderators are unknown is version (3), in which we include all moderators at once.

In the nodewise regression algorithm used to estimate all three versions of the moderated network models, we select the tuning parameters $\lambda_s$ that minimizes the Extended Bayesian Information Criterion (EBIC), which has been shown to perform well in recovering sparse graphs \citep{foygel2010extended}. The EBIC is an extension of the BIC \citep{schwarz1978estimating} in that it puts an additional penalty on the number of nonzero parameters. This additional penality is weighted by a parameter $\gamma$. We set $\gamma = 0.5$ because this value led to good performance in simulations using a setting similar to ours \citep{2016arXiv160605771E}.

\subsubsection{Network Comparison Test (NCT)}\label{sec_sim_est_NCT}

The  Network Comparison Test (NCT) performs a permutation test for each edge parameter to determine whether it is reliably different across two groups (data sets). Since the NCT makes a comparison between two groups it requires to split the dataset in half. Here we split at the median of the moderator variable. In a first step, the NCT estimates a model on each data set and takes the absolute value the differences between the corresponding parameters. These differences serve as test statistics. In a second step, a sampling distribution under the null hypothesis (no difference) is created for each edge comparison by $B$ times randomly permuting the group membership of data points, estimating the two models and computing the absolute value of all edge differences. This gives sampling distributions for each edge-parameter difference, which can be used to test the significance of the edge-difference from step 1 under the null hypothesis that there is no difference. For estimation, the NCT uses the graphical lasso algorithm \citep{friedman2008sparse} and selects the regularization parameter $\lambda$ with using the EBIC with hyperparameter $\gamma = 0.5$. We used $B = 1000$ and set the significance threshold to $\alpha = 0.05$. In our simulation, for small sample sizes $n=30, 46$ the covariance matrices for the two groups (each computed from $n=\frac{30}{2}, \frac{46}{2}$) were not positive definite for some of the $B$ bootstrap samples. To still be able to run the NCT algorithm, we modified the original algorithm by \cite{van2016comparing} in that we project these covariance matrices to the nearest positive definite covariance matrix\footnote{We use the implementation of the algorithm of \cite{higham2002computing} in the R-package \emph{Matrix} \citep{Matrixpackage}.}.

\subsubsection{The Fused Group Lasso (FGL)}\label{sec_sim_est_FGL}

The Fused Group Lasso (FGL) \citep{danaher2014joint, costantini2017stability} jointly estimates two GGMs by using two separate $\ell_1$-penalties: the first penalty (weighted by $\lambda_1$) includes all parameters of the model (two covariance matrices), which is the standard $\ell_1$-penalty \citep[e.g.,][]{hastie2015statistical}. This penalty is similar to the one we use to estimate the moderated network models; the second penalty (weighted by $\lambda_2$) includes the \emph{difference} of the two covariance matrices, and therefore penalizes parameter differences across groups. For a detailed description of the FGL see \cite{danaher2014joint}. Following the implementation in the R-package \emph{EstimateGroupNetwork} \citep{FGLpackage}, we first select $\lambda_1$ and then $\lambda_2$, using the EBIC with $\gamma = 0.5$. We do not perform a full grid-search on $\lambda_1, \lambda_2$ since this is computationally very expensive \citep{costantini2017stability}. Because the FGL jointly estimates two GGMs on two data sets, also here we median-split the data set along the moderator variable.

For both the NCT and the FGL methods we run two versions: (1) we create two groups by splitting the dataset at the median value of the moderator variable and then run the NCT/FGL; (2) we create the grouping for each of the $p$ variables in the data set one by one and compute the NCT/FGL for each of those groupings. And then take the union of detected moderation effects as output. Since no more group differences can be discovered in the additional $p-1=12$ runs, the sensitivity cannot improve. Again, we include this condition to investigate the drop of precision when using such an exploratory approach. The estimates of NCT/FGL can be seen as a piecewise constant approximation to the linear moderation effect, with two constant functions on the left/right of the median split on the moderator variable.

We estimate moderated network models using the implementation in the R-package \emph{mgm} \citep{haslbeck2017mgm}. The NCT is implemented in the R-package \emph{NetworkComparisonTest} \citep{NCTpackage} and the FGL is implemented in the R-package \emph{EstimateGroupNetwork} \citep{FGLpackage}. All three packages are open source and freely available on the Comprehensive R Archive Network (CRAN) (\url{https://cran.r-project.org/}).

\subsection{Results}\label{sec_sim_results}

We report \emph{sensitivity} (probability of recovering a true parameter) and \emph{precision} (probability that an estimated parameter is a true parameter). For the moderated network models recovering a true (nonzero) parameter means estimating a nonzero parameter with positive sign. For NCT and FGL, recovering a moderation effect means that the group difference of a given pairwise interaction is significant (NCT) or nonzero (FGL) and that the parameter estimated on the data set with larger values on the moderator variable is larger (that is, the difference has the correct sign). In the FGL version (2), which runs over all $\{1, 2, \dots, p \}$ possible moderators, it can happen that a given edge difference is detected with different signs in several runs. We then select the difference with the largest absolute value. 

\begin{figure}[H]
	\centering
	\includegraphics[width=1\textwidth]{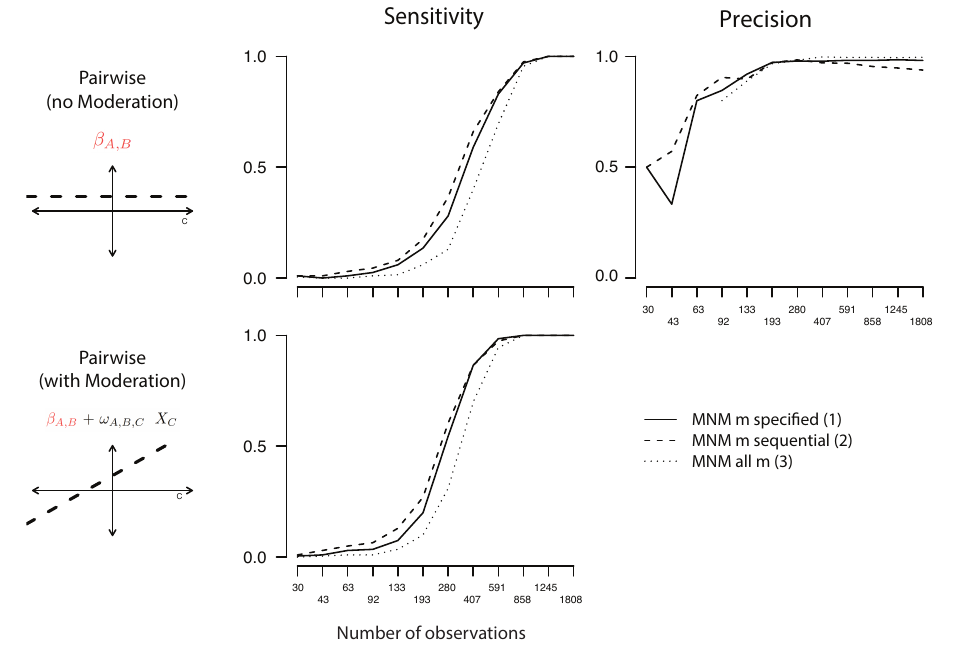}
	\caption{Sensitivity and precision for the moderated network model estimators for pairwise interactions without moderation (first row) and pairwise interactions with moderation (second row), as a function of $n$ on a log scale. Precision is defined with respect to all pairwise estimates and is therefore the same for pairwise parameters with/without moderation.}\label{fig_sim_results_part1}
\end{figure}

Figure \ref{fig_sim_results_part1} shows the average sensitivity and precision of pairwise interactions and the pairwise part in a partially moderated pairwise interaction over 100 iterations. Precision is defined with respect to \emph{all} pairwise interactions (moderated or not) and is therefore only displayed once. Note that the sensitivity for small $n$ is very low. This implies that precision is undefined in many iterations. We display precision only if precision is defined in at least 5 iterations. We report the performance for the three estimation versions of the moderated network model.

We first turn to the performance of the $\ell_1$-regularized nodewise regression in estimating the pairwise parameters in the MNM. The sensitivity of all three versions of the moderated network model seems to converge to 1 when increasing $n$. The different versions of the network model stack up as expected: the exploratory sequential version (2) has the highest sensitivity, since combining the standard version with specified moderator (1) with the estimates of $p-1$ additional models can only increase the sensitivity. The network model with all possible moderator effects specified at once has the lowest sensitivity. This makes sense, since it has a much larger number of parameters and hence larger regularization parameters $\lambda_s$ to control the variance of the estimates. Consequently, precision stacks up in reverse order. The precision for versions (1) and (3) seems to converge to 1, while the precision of (2) does not. 

\begin{figure}[H]
	\centering
	\includegraphics[width=1\textwidth]{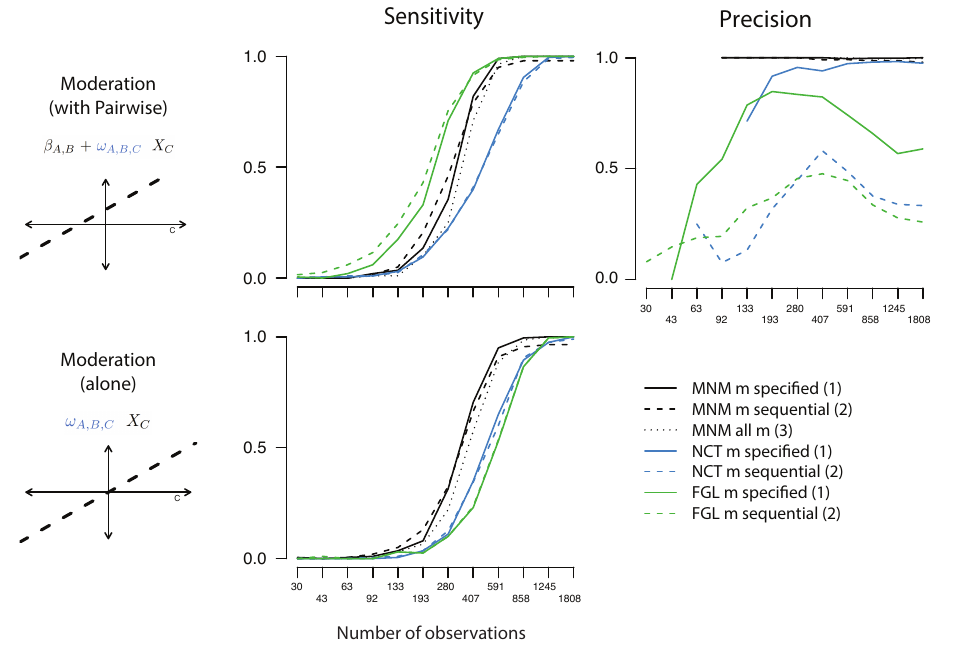}
	\caption{Sensitivity and precision for the moderated network model estimators and the NCT and FGL estimators for the moderation effects in a partially moderated pairwise interaction (first row) and moderation effects in a fully moderated pairwise interaction (second row), as a function of $n$ on a log scale.}\label{fig_sim_results_part2}
\end{figure}

Figure \ref{fig_sim_results_part2} shows that all of the results described in the previous paragraph also hold for moderation effects. The performance to recover pairwise and moderation effects is similar. The largest difference in performance between parameter types is between the sensitivity to detect the unmoderated pairwise interaction (row 1 in Figure \ref{fig_sim_results_part1}) and the full moderation (row 2 in Figure \ref{fig_sim_results_part2}) is \textit{smaller} compared to the sensitivity to detect the pairwise and moderation effects in the partially moderated pairwise interaction (row 2 in Figure \ref{fig_sim_results_part1} and row 1 in Figure \ref{fig_sim_results_part2}, respectively). 

Turning to the NCT, Figure \ref{fig_sim_results_part2} shows that its sensitivity seems to converge to 1, but does so slower than all other methods for partial moderation (first row), and comparable to the FGL for full moderation effects (second row). Precision grows slower than for MNMs and is only close to 1 for $n \geq 591$ observations. The version of the NCT that searches for all $p$ possible moderators sequentially cannot improve sensitivity, because there is nothing to detect if an incorrect moderator is specified. The precision of the sequential NCT is low and does not converge to 1 as $n$ increases.

The FGL shows the highest sensitivity in recovering moderation effects in a pairwise interaction (row 1) and shows the lowest sensitivity (comparable to NCT) for full moderation (row 2). Within FGL, the sequential version shows higher sensitivity while above we claimed that the sensitivity cannot be larger than in the specified version. The precision of the FGL with specified moderator increases up to $n = 193$ and decreases for larger $n$. The sequential FGL has low precision for all $n$. 

In Appendix \ref{app_results_in_tables} we also provide the results shown in Figures \ref{fig_sim_results_part1} and \ref{fig_sim_results_part2} in tables.

\subsection{Discussion of Simulation Results}\label{sec_sim_discussion}

The goal of this simulation was (a) to investigate the performance of $\ell_1$-regularized nodewise regression for estimating moderated network models and (b) compare its performance to detect moderation effects to the split-sample methods NCT and FGL.

\subsubsection{Performance of Moderated Network Models}

The MNMs with correctly specified moderator (version 1) and with all moderators specified at once (version 3) are consistent estimators for moderated network models in the setting of our simulation, since both their sensitivity and precision seem to converge to 1 as $n$ increases. Version 2, which combines results of $p$ sequential moderated network models showed similar performance, but does not converge to 1 for the sequence of $n$ investigated in our simulation. The reason is that false-positives accumulate across the $p$ models. When considering larger $n$ we would expect that also version 2 converges.

The second important finding is that pairwise interactions and moderation effects (3-way interactions) are roughly equally difficult to estimate. From a $\ell_1$-regularized nodewise (LASSO) regression perspective this is what we would expect: moderation effects are just additional predictors that are uncorrelated with the respective main effects (we show that in Appendix \ref{app_centering}). For estimation, moderation effects are therefore in no relevant way different from main effects and hence a different performance in estimating them would be surprising. The fact that moderation effects are just additional predictors in a regularized regression means that we can make use of the large pool of theoretical and simulation results on the performance of $\ell_1$-regularized regression in different situations \citep[e.g.][]{hastie2015statistical, buhlmann2011statistics}. Theoretical results of the LASSO for nodewise regression require the assumption of sparsity. Graph-sparsity does not apply to moderated network models since their parameters cannot be represented in a $p \times p$ graph (see Section \ref{sec_intro_viz_modNW}). 

Based on anecdotal evidence we expect that in reality moderation effects are on average smaller than pairwise interactions. If this is true, it will be harder to recover any moderation effects than recovering any pairwise effects in the trivial sense that smaller effects are harder to recover than larger effects. In the simulation study we kept the size of pairwise interactions and moderation effects equal to investigate the presence of any unexpected effects. This was important so as to verify whether we can use theory on multiple regression to make predictions about the performance of moderated network models in different situations (we can).

The largest performance difference across the four parameter types was between the sensitivity of unmoderated pairwise interaction and full moderation on the one hand, and the pairwise and moderation effects in the partially moderated interaction on the other hand. This difference is explained by the fact that in the latter cases the number of uncorrelated predictors is larger, which leads to a small increase in sensitivity. In Appendix \ref{app_results_compareTypes} we provide a figure that directly displays this difference. In Appendices \ref{app_basicSim} and \ref{app_results_neighbors} we show with additional simulations that the above explanation is correct. Specifically, we show that this phenomenon is also present in unmoderated network models and thus not specific to MNMs.

In the present simulation we specified the MNM to include only a single moderator variable. We chose to include a single moderator variable to ensure that the simulation setup is easy to understand and to keep the sampling procedure feasible. However, the performance of the estimators with all moderators specified, or all moderators specified sequentially will not change much when including an additional moderator. The reason is that all moderators are already included in the model. Including one additional parameter therefore does not change the estimated model, it merely means that one additional parameter is nonzero in the true MNM. Since we estimate the MNM nodewise, and because higher- and lower-order terms are uncorrelated (see Appendix \ref{app_centering}), adding a moderator in the present situation is similar to fitting a fixed linear regression model and setting one parameter in the true model from zero to nonzero.

\subsubsection{Performance of NCT and FGL}

Sensitivity and precision of the NCT in which we specified the correct moderator seem to converge to 1 as a function of $n$. However, both sensitivity and precision are lower than for the moderated network model in all situations included in our simulation. The sequential version of the NCT has low precision for all $n$ and should therefore not be used in situations similar to the one used in our simulation.

Turning to the FGL version in which we also specified the correct moderator, its sensitivity is comparable to the NCT for full moderation and largest across all methods for moderation in partially moderated interactions. The explanation for this difference in sensitivity is a combination of two factors: first, the EBIC selects models with higher sensitivity if the number of uncorrelated neighbors is larger (we show this in Appendix \ref{app_results_neighbors}). In the moderation with pairwise effect (row 1, Figure \ref{fig_sim_results_part2}) there is an additional predictor (neighbor) compared to the moderation effect alone (row 2). Second, the graphical lasso is more liberal than nodewise regression. The second factor combined with the first one explains why the sensitivity difference is larger for the FGL compared to the moderated network models. However, since the FGL is not the focus of the present work we did not further investigate this difference. The difference in sensitivity between the standard and the sequential version of the FGL is explained by the fact that by design 4-8 variables are correlated with the moderator 13 (see Figure \ref{fig_sim_generating}). Therefore, splitting by these variables in the sequential version is similar to splitting along the moderator proportional to the correlation between the moderator and the splitting variable. The precision of the FGL with specified moderator increases up to $n = 193$ and decreases for larger $n$. The sequential FGL has low precision for all $n$. A partial explanation for this low precision of the FGL could be the low precision of the graphical lasso algorithm for large $n$ if its assumption of sparsity if violated \citep{williams2018back}. Because of its low precision, we do not recommend the FGL for situations similar to the one in our simulation.

So far, we only discussed the version of the NCT and FGL in which the true moderator variable was known and provided to the algorithms. However, this may not be the case in many realistic situations. In version (2) of NCT and FGL we ran the algorithms $p$ times, each time specifying another variable as the moderator and combining the results (for details see above). Figure \ref{fig_sim_results_part2} shows that the sensitivity of this approach is (as expected) a bit higher, however, precision is very low.

\subsubsection{Moderated Network Models vs. Sample-split Methods}

For the situations considered in our simulation study, we found that moderated network models are consistent estimators for moderation effects and outperform the split-sample based methods NCT and FGL. We generated the data from a MNM and therefore our model is correctly specified, while the sample-split methods NCT/FGL only approximate the linear moderation effect with a piecewise constant function. We therefore expected that our model would perform better. If the moderator is a Bernoulli random variable and NCT/FGL do not have the disadvantage of only approximating the moderation effect, the performance difference may be smaller. However, this comparison would require moderated MGMs, which we leave for future work (see also Discussion in Section \ref{sec_Discussion}). Another important \emph{general} advantage of the moderated network approach is that it has much larger sensitivity to detect pairwise interactions, because the method does not require to split the sample in half.

\section{Empirical Data Examples}\label{sec_dataexamples}

In this section, we apply Moderated Network Models to empirical data. Specifically, we provide a fully reproducible tutorial on how to fit MNMs to a data set of mood variables using the R-package \emph{mgm} \citep{haslbeck2017mgm} (Section \ref{sec_tutorial}). We also present different options for visualizing MNMs using factor graphs (see Section \ref{sec_intro_viz_modNW}). On the basis of the same data set we then discuss possible aspects of model misspecification and tools to detect those (Section \ref{sec_misspec}).

\subsection{R-Tutorial: Fit Moderated Network Model to Data Set of Mood Variables}\label{sec_tutorial}

We show how to fit a MNM to a cross-sectional data set consisting of $n = 3896$ observations of the five mood variables \emph{hostile}, \emph{lonely}, \emph{nervous}, \emph{sleepy} and \emph{depressed} with 5 response categories. This data set is a subset of the data set \verb|msq| from the R-package \emph{psych} \citep{R_psych}. To fit the MNM, we use the R-package \emph{mgm}, which implements functions to estimate $k$-order Mixed Graphical Models (MGMs), of which GGMs and MNMs are special cases. The package can be installed and loaded in the following way:

\begin{verbatim}
install.packages("mgm")
library(mgm)
\end{verbatim}

In the following two subsections we first show how to fit the MNM to the data and then present possible visualizations of the model.

\subsubsection{Fit Moderated Network Model to Data}

The data set containing the five trait mean scores is automatically loaded with \emph{mgm} and available as the object \verb|msq_p5|:

\begin{verbatim}
> dim(msq_p5)
[1] 3896    5
> head(msq_p5)
     hostile     lonely    nervous     sleepy  depressed
1 -0.4879522  0.7280507  1.0018084 -0.2334325 -0.5998642
2 -0.4879522 -0.6442210 -0.5445646 -0.2334325 -0.5998642
3 -0.4879522 -0.6442210 -0.5445646 -1.1857391 -0.5998642
4 -0.4879522  0.7280507  2.5481814 -0.2334325  0.8672236
5 -0.4879522 -0.6442210 -0.5445646 -0.2334325 -0.5998642
6 -0.4879522 -0.6442210 -0.5445646  0.7188742  0.8672236
\end{verbatim}

\noindent
\verb|dim(msq_p5)| shows that the dataset consists of 3896 rows and 5 columns and \verb|head(msq_p5)| displays the first 6 rows of the data set. The data points have several points after the decimal because each variable was scaled to mean zero and a standard deviation of one.

We provide the data in \verb|msq_p5| to the estimation function \verb|mgm()| of the \emph{mgm} package.  Next to the data, we specify the types and levels for each variable. Since we model all variables as Gaussian distributions, we specify \verb|"g"| for each variable and the number of levels as \verb|1| by convention for continuous variables. This specification is necessary in \emph{mgm}, because the package also allows to model Poisson variables and categorical variables with $m$ categories. Via the argument \verb|moderator| one specifies the moderators to be included in the model. For instance, if we select \verb|moderators = c(1, 3)| all moderation effects of variables 1 and 3 are included in the model. Here we do not have any prior theory about possible moderators and therefore specify all variables as moderators by setting \verb|moderators = 1:5|. This corresponds to version 3 of the estimator for MNMs in the simulation study (Section \ref{sec_simulation}).

The estimation algorithm uses $p=5$ nodewise penalized regressions, for each of which an appropriate regularization parameter $\lambda_s$ has to be selected (see Section \ref{methods_nodewisereg}). We select the $\lambda_s$ that minimizes the EBIC with the hyperparameter $\gamma = 0.5$ by setting \verb|lambdaSel = "EBIC"| and  \verb|lambdaGam = .5|. This is the same setup we used in the simulation study. Alternatively one could select $\lambda_s$ using cross-validation (\verb|lambdaSel = "CV"|). With \verb|scale = TRUE| we specify that all predictors are scaled to mean zero and standard deviation 1. This is a standard procedure in regularized regression and avoids that the penalization of a given parameter depends on the standard deviation of the associated variable (see Section \ref{methods_nodewisereg}). With \verb|ruleReg = "AND"| we specify that the nodewise regressions are combined with the AND-rule (see Section \ref{methods_nodewisereg}).

\begin{verbatim}
mgm_mod <- mgm(data = msq_p5,
               type = rep("g", 5),
               level = rep(1, 5),
               lambdaSel = "EBIC",
               lambdaGam = .5,
               ruleReg = "AND",
               moderators = 1:5,
               scale = TRUE)
\end{verbatim}

The main output is stored in \verb|mgm_mod$interactions|. For a detailed description of the output see the help file \verb|?mgm| and the \emph{mgm} paper \citep{haslbeck2017mgm}. The list entry \verb|mgm_mod$interactions$indicator| contains a list of all estimated parameters separately for each order (2-way, 3-way, etc.):

\begin{verbatim}
> mgm_mod$interactions$indicator
[[1]]
[,1] [,2]
[1,]    1    3
[2,]    1    4
[3,]    1    5
[4,]    2    3
[5,]    2    4
[6,]    2    5
[7,]    3    4
[8,]    3    5
[9,]    4    5

[[2]]
[,1] [,2] [,3]
[1,]    1    2    3
[2,]    1    2    4
[3,]    1    3    4
[4,]    3    4    5
\end{verbatim}

The first level of this list shows that there are nine pairwise interactions; and the second entry shows that there are four moderation effects (or 3-way interactions). Specifically, the entry \verb|mgm_mod$interactions$indicator[[1]][6, ]| indicates that there is a nonzero pairwise interaction between variables 2-5. And the entry \verb|mgm_mod$interactions$indicator[[2]][4, ]| indicates that there is a nonzero moderation effect between variables 3-4-5. To obtain more information about a given interaction we use the function \verb|showInteraction()|. One can obtain the parameter for the pairwise interaction 2-5 (\emph{lonely} and \emph{depressed}) via:

\begin{verbatim}
> showInteraction(object = mgm_mod, int = c(2,5))
Interaction: 2-5 
Weight:  0.4318148 
Sign:  1  (Positive)
\end{verbatim}

This pairwise interaction can be interpreted as in a linear regression: when increasing \emph{lonely} by one unit, \emph{depressed} increases by $\approx 0.432$ units, when keeping all other variables constant. The parameters for the moderation effect can be obtained similarly: to obtain the moderation effect between \emph{nervous}, \emph{sleepy} and \emph{depressed} we provide the respective column numbers to the \verb|int| argument:

\begin{verbatim}
> showInteraction(object = mgm_mod, int = c(3,4,5))
Interaction: 3-4-5 
Weight:  0.0564465 
Sign:  1  (Positive)
\end{verbatim}

We can interpret this moderation effect in three different ways: First, he pairwise interaction between \emph{nervous} and \emph{sleepy} is equal to zero when \emph{depressed} is equal to zero (no pairwise interaction between \emph{nervous} and \emph{sleepy}) and increases by $\approx 0.06$ when increasing \emph{depressed} by one unit. For the other two interpretations we need the parameters of the pairwise interactions between \emph{depressed} and \emph{sleepy}, and between \emph{depressed} and \emph{nervous}:

\begin{verbatim}
> showInteraction(object = mgm_mod, int = c(4,5))
Interaction: 4-5 
Weight:  0.1534387 
Sign:  1  (Positive)> 
> showInteraction(object = mgm_mod, int = c(3,5))
Interaction: 3-5 
Weight:  0.1029161 
Sign:  1  (Positive)
\end{verbatim}

The second interpretation is: the pairwise interaction between \emph{depressed} and \emph{sleepy} is equal to $\approx 0.153$ when \emph{nervous} is equal to zero, and increases by  $\approx 0.06$ when increasing \emph{nervous} by one unit. Similarly, the third interpretation is that the pairwise interaction between \emph{depressed} and \emph{nervous} is equal to $\approx 0.103$, when \emph{sleepy} is equal to zero, and increases by $\approx 0.06$ when increasing \emph{sleepy} by one unit. For example, if \emph{sleepy} has the value $2$, then the pairwise interaction parameter between \emph{depressed} and \emph{nervous} is equal to  $\approx  0.103 + 2 \times 0.06 = 0.223$

\subsubsection{Visualize Moderated Network Model as Factor Graph}

In many situations it is more convenient to inspect the model parameters graphically. Since the MNM contains more than $p \times p$ parameters, they cannot be visualized in a standard network with $p \times p$ edges. Instead, we use a factor graph, in which we introduce additional nodes for interactions (for details see Section  \ref{sec_intro_viz_modNW}). The function \verb|FactorGraph()| from the \emph{mgm} package draws such factor graphs from the output objects of \verb|mgm()|:

\begin{verbatim}
FactorGraph(object = mgm_mod,
            edge.labels = TRUE, 
            labels = colnames(msq_p5))
\end{verbatim}

With \verb|edge.labels = TRUE| we specified that the values of parameters are shown the visualization. Figure \ref{fig_appl_factorgraph} (a) shows the resulting plot:

\begin{figure}[H]
	\centering
	\includegraphics[width=1\textwidth]{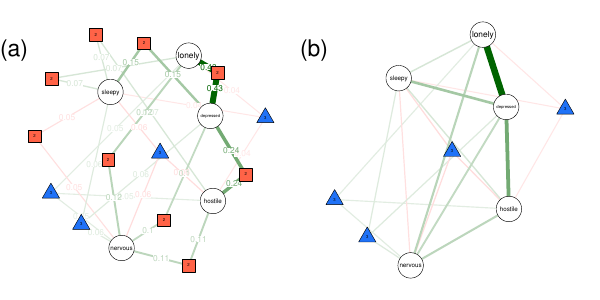}
	\caption{Two different factor graph visualizations: (a) variable-nodes are displayed as circle nodes, pairwise interactions are displayed as square nodes, and moderation effects (3-way interactions) are displayed as triangles; (b) Only moderation effects (3-way interactions) are displayed as triangle nodes, pairwise interactions are displayed as simple edges. Green edges indicate parameters with positive sign. The widths of edges is proportional to the absolute value of the parameter. }\label{fig_appl_factorgraph}
\end{figure}

The green (red) edges indicate parameters with positive (negative) sign, and the width of edges is proportional to the absolute value of the parameter. If two variables are connected by a pairwise interaction but not by a 3-way interaction, the pairwise interaction parameter is a partial correlation. If two variables are connected by a pairwise interaction \emph{and} and 3-way interaction, the pairwise interaction parameter is the partial correlation between the two variables if the third variable (in the 3-way interaction) is equal to zero (see Section \ref{sec_methods_simplemoderation}). The function \verb|FactorGraph()| is a wrapper around the \verb|qgraph()| function from the \emph{qgraph} package \citep{epskamp2012qgraph} and all \verb|qgraph()| arguments can passed to customize the visualization.

For models that include many pairwise interactions this visualization may become unclear. To alleviate this problem, the factor nodes representing pairwise interactions can be replaced by simple edges with \verb|PairwiseAsEdge = TRUE|. In addition, we specified with  \verb|edge.labels = FALSE| that the parameter values are not shown. The resulting visualization is shown in Figure \ref{fig_appl_factorgraph} (b). While this graph is not a traditional factor graph anymore, the visualization contains the same information as the visualization in (a), except the parameter values.

An alternative way to visualize MNMs is to condition on a set of values of the nonzero moderators and visualize the resulting pairwise network. This can be done with the function \verb|condition()| which takes the model object \verb|mgm_mod| and a list assigning a value to each moderator as input. The function outputs a new, conditioned, pairwise model object that can be visualized as a network. Repeating this process for a number of values of the nonzero moderators allows to show the pairwise network as a function of the nonzero moderators. This is especially useful if there is only a single moderator in the model. In the case of a large number of moderators, this approach becomes unfeasible, because the number of values to map out the space of moderator variables (and therefore the number of networks to plot) becomes grows exponentially with the number of moderators. 

In Appendix \ref{app_simdata_tutorial} we provide an additional tutorial in which we recover the MNM used in iteration 2 of the simulation study in Section \ref{sec_simulation}.

\subsection{Model Misspecification}\label{sec_misspec}

Like for any other statistical model, when fitting MNMs to empirical data we assume that these data were generated by the class of MNMs, that is, we assume that the model is correctly specified. Specifically, we constructed the MNM from $p$ conditional Gaussian distributions and therefore assume that each variable is conditionally Gaussian. In addition, we assume that the mean of each variables is modeled by a regression equation of the form of equation (\ref{eq_conditional_mean}). 

\subsubsection{Types of Model Misspecification}

The MNM is estimated via $p$ moderated multiple regression models and therefore the possible types of model misspecification are the same as in multiple regression with moderation/interactions. The first type of misspecification in these models is the presence of non-linear effects (in pairwise or/and moderation effects); the second type is the presence of conditional distributions that are not Gaussian distributed. Both types of problems are well documented in the regression literature (e.g. \cite{aiken1991multiple, afshartous2011key, hainmueller2018much}), which is why we do not discuss them here in detail.

Instead, we focus on a new type of misspecification that arises from constructing the MNM joint distribution from the $p$ conditional Gaussian distributions. Specifically, our construction of the MNW implies that the population moderation effect of $A$ on the pairwise interaction $B, C$ is the same as the moderation effect of $B$ on $A, C$ and $C$ on $A, B$ (and the three parameters converge empirically as $n \rightarrow \infty$). It is this equality that justifies aggregating these three parameter estimates using the AND- or OR-rule (see Section \ref{methods_nodewisereg}). If all variables are generated from a joint distribution that can be factorized into conditional Gaussians, the moderation effects are the same across conditional distributions (nodewise regressions). However, if the data are skewed, it is possible that the moderation effects are \emph{different} across nodewise regressions and summarizing them in a single parameter would be misleading. This does not mean that the moderation effects always differ across nodewise regressions if the data is skewed. For example, the data used in the previous section were skewed but the moderation effects were roughly the same across nodewise regressions. In the following subsection we give an empirical data example with skewed variables in which moderation effects \emph{are} different across nodewise regressions.

\subsubsection{Different Moderation Effects across Nodewise Regressions}\label{sec_misspec_signs}

We illustrate the problem of largely varying moderation effects across nodewise regressions using a data set consisting of $n = 3896$ observations of the mood variables \emph{afraid}, \emph{ashamed} and \emph{distressed}. These data are also from the data set \verb|msq| from the R-package \emph{psych} \citep{R_psych} and are loaded with the \emph{mgm} package as the object \verb|msq_p3|. We request the dimensions and the first six rows of the data set with the following code:

\begin{verbatim}
> dim(msq_p3)
[1] 3896    3
> head(msq_p3)
      afraid   ashamed distressed
1  2.1854037 -0.280891  0.8504921
2 -0.2853835 -0.280891 -0.5565216
3 -0.2853835 -0.280891 -0.5565216
4 -0.2853835  2.067502  0.8504921
5 -0.2853835 -0.280891 -0.5565216
6 -0.2853835 -0.280891 -0.5565216
\end{verbatim}

The histograms in Figure \ref{fig_misspec_psych_hists} show that all three variables are highly skewed:

\begin{figure}[H]
	\centering
	\includegraphics[width=1\textwidth]{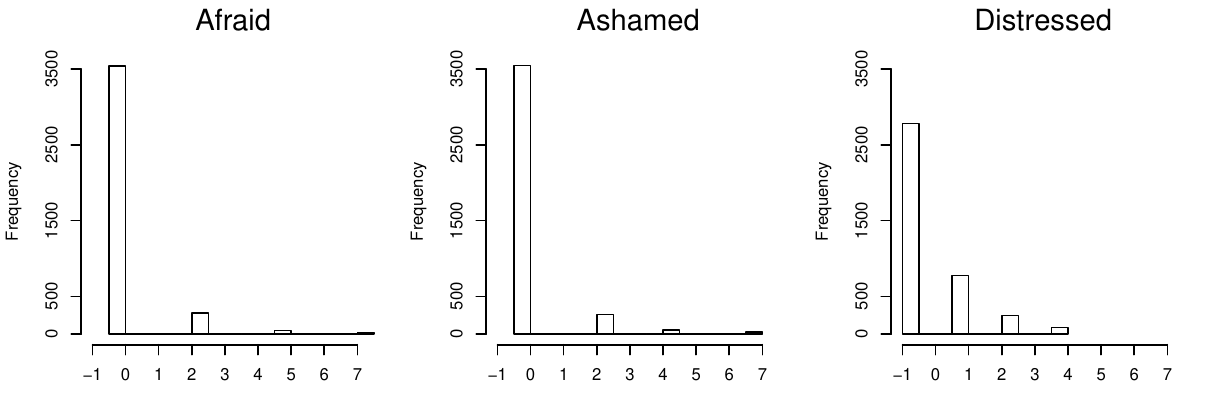}
	\caption{Histograms for the scaled variables \emph{afraid},  \emph{ashamed} and \emph{distressed}.}\label{fig_misspec_psych_hists}
\end{figure}

In Appendix \ref{app_misspec_modeffects}, we show conditional scatter plots to provide some intuition for how it is possible that moderation effects differ across nodewise regressions. Here we proceed by estimating a moderated network model with the same specifications as above in Section \ref{sec_tutorial}:

\begin{verbatim}
mgm_mod2 <- mgm(data = msq_p3,
                type = rep("g", 3),
                level = rep(1, 3),
                lambdaSel = "EBIC",
                lambdaGam = .5,
                ruleReg = "AND",
                moderators = 1:3)
\end{verbatim}

Again similarly to above we use the function \verb|FactorGraph()| to plot the factor graph visualization of the moderated network model

\begin{verbatim}
FactorGraph(object = mgm_mod2, 
            edge.labels = TRUE)
\end{verbatim}

\noindent
which is shown in Figure \ref{fig_misspec_msq_p3_factorgraphs} (a):

\begin{figure}[H]
	\centering
	\includegraphics[width=1\textwidth]{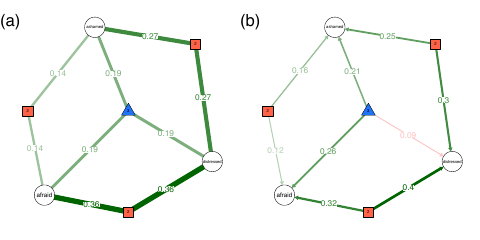}
	\caption{(a) Factor graph visualization of the estimated network model with aggregated parameter values; (b) Nodewise factor graph visualization of all estimated nodewise (unaggregated) parameters. The direction of edges indicates the nodewise regression in which the parameter has been estimated.}\label{fig_misspec_msq_p3_factorgraphs}
\end{figure}

Panel (a) in Figure \ref{fig_misspec_msq_p3_factorgraphs} shows that there is a moderation effect with value $\approx 0.19$. We can interpret this moderation effect in the following way: the pairwise interaction between \emph{distressed} and \emph{ashamed} is equal to $\approx 0.27$ when \emph{afraid} is equal to zero, and increases by $\approx 0.19$ when increasing \emph{afraid} by one unit. However, we now show that this interpretation is inappropriate, because the nodewise estimates differ widely and the aggregate parameter is therefore misleading.

The nodewise parameter estimates can be accessed via the \verb|mgm()| output object (see \verb|?mgm|). A more convenient way to inspect the unaggregated nodewise estimates is to plot them into a directed version of the factor graph, in which the direction of the edges indicates the nodewise regression in which the parameter has been estimated. This modified Factor graph can be plotted by setting \verb|Nodewise = TRUE| in the  \verb|FactorGraph()| function:

\begin{verbatim}
FactorGraph(object = mgm_mod2, 
            edge.labels = TRUE, 
            Nodewise = TRUE)
\end{verbatim}

Figure \ref{fig_misspec_msq_p3_factorgraphs} (b) shows the resulting visualization. A directed edge towards a given node always indicates a parameter obtained from the regression on that node. For example, the directed edge towards \emph{afraid} from the order-2 factor node that connects \emph{afraid} with \emph{distressed} indicates the pairwise interaction between \emph{afraid} and \emph{distressed} obtained from the nodewise regression on \emph{afraid}. Importantly, the directionality of the edges shown in Figure \ref{fig_misspec_msq_p3_factorgraphs} (b) only stems from the nodewise regression algorithm and does not represent an actual directionality of the effect.

We see that the moderation effect on \emph{distressed} is actually negative. The correct interpretation would therefore be that the pairwise interaction between \emph{distressed} and \emph{ashamed} is equal to $\approx 0.27$ when \emph{afraid} is equal to zero, and \emph{decreases} by $\approx 0.09$ when increasing \emph{afraid} by one unit. This is a moderation effect in the opposite direction of the aggregate moderation effect we used in the interpretation above.

What should one do in such a situation? There is no easy answer. Clearly, interpreting the parameters of the moderated network model joint distribution shown in (a) is misleading and therefore no option. A solution would be to reject the joint distribution in (a) and instead report the combined conditional distributions in (b). This has the downside that the model is more complex and that the joint distribution is unknown. The latter is undesirable, because this means that we cannot perform inference on the joint distribution. On the other hand, in many applications of network analysis in psychology no such inference is performed. The principled solution would be to create a joint distribution that incorporates skewed distributions and moderation effects that vary across conditional distributions. However, we expect this to be difficult and far beyond the scope of the present paper.

\section{Discussion}\label{sec_Discussion}

We introduced Moderated Network Models by using the standard definition of moderation in the regression framework and adding moderation effects to the multivariate Gaussian distribution. We presented a new visualization for Moderated Network Models based on factor graphs and we proposed an $\ell_1$-regularized nodewise regression procedure to estimate this model. In a simulation study we reported the performance of this approach to recover different types of parameters in a random graph with moderation and showed that estimating a moderated network model outperforms the split-sample based methods Fused Graphical Lasso (FGL) and the Network Comparison Test (NCT). Finally, we provided a fully reproducible tutorial on how to estimate MNMs with the R-package \emph{mgm} and discuss possible issues with model misspecification.

Three limitations are important to keep in mind. First, as discussed in Section \ref{sec_construction}, we do not have an explicit constraint on the parameter space of the MNM joint distribution that ensures that the distribution is normalizable. In order to sample observations, we worked around this issue with a rejection sampler (see Appendix \ref{app_Sampling}). But for a given model estimated from data, it is unclear whether the resulting joint distribution is normalizable. This means that while the joint distribution does capture the dependency structure in the data, it might not be possible to define a probability distribution over it. A consequence of the absence of a guaranteed probability distribution is that one cannot use global likelihood ratio tests or goodness of fit analyses to select between models. We expect that an appropriate constraint on the parameter space to be difficult to work out since it involves all parameters of the model and the variances of all conditional Gaussians. While this is an important limitation to keep in mind, all conditional distributions \emph{are} consistently estimated proper distributions. Thus, if inferences are limited to the conditional distribution there is no issue. This implies that predictions for any variable can be computed without any limitation, which allows to perform model selection using out of sample prediction error. Also, it is important to keep in mind that the joint distribution correctly captures the dependency structure. This implies that one can use network statistics such as centrality metrics, modularity or global efficiency to describe the global network structure.

Second, the performance results obtained from our simulation study may be different in other setups. The best way to obtain the performance in setups with larger/smaller parameters or higher/lower sparsity is to run appropriate additional simulation studies. That said, recall that we estimate the MNM using a series of multiple regression models with interaction terms. Since interaction effects can be seen as additional variables in a multiple regression, the performance in recovering is in principle the same as for main effects. This means that one can draw on the rich literature on these models to make predictions about the performance in recovering MNMs in different setups \citep[e.g.][]{hastie2015statistical, buhlmann2011statistics}.

Third, in our simulation study we assigned the same size to pairwise interactions and moderation effects. We did this to confirm our prediction from the linear regression framework that pairwise and moderation effects are equally hard to estimate. But in reality moderation effects are often much smaller than pairwise interactions. This means that moderation effects are on average harder to detect than pairwise interactions in the same sense as small pairwise interactions are harder to detect than large pairwise interactions. It is possible to make the estimation procedure more liberal by reducing the hyperparameter $\gamma$ in the EBIC to detect small moderation effects. However, this would result in higher false positive rates for both pairwise and moderation effects. In sum, moderation effects tend to be smaller than pairwise interaction and are therefore harder to detect by \emph{any} estimation algorithm. However, MNMs will recover moderation effects that have a size comparable to pairwise interactions. And any moderation effect can be recovered if enough data is available.

We see the following extensions for future research. First, in this paper we suggested a $\ell_1$-regularized nodewise regression approach to estimate moderated network models. We chose this estimator because it deals well with the large number of parameters and renders the interpretation of the model easier by setting small parameters to zero. In addition, the underlying assumption that most parameters are zero is reasonable in the case of MNMs (see Section \ref{methods_nodewisereg}). But in some situations a different estimator might perform better. One example is the presence of correlated predictors in skewed data. In this situation, the lower-order terms will be correlated with the higher-order terms in the regression models (see Section \ref{app_misspec_modeffects} for an illustration of that fact). If this correlation becomes too large, the $\ell_1$-regularized estimator selects only one of the two terms. This problem is larger, if $n$ is small and the selected regularization parameter $\lambda$ large. It would be interesting to map out how problematic this is and whether other estimators, for example based on significance testing, are better suited for such situations.

Second, the MNM we propose includes all pairwise interactions and a set of specified 3-way interactions/moderation effects. Since we freely estimate all specified parameters, it is possible that a moderation effect is estimated to be nonzero, while none of the pairwise interactions between the involved variables is estimated to be nonzero. In some situations, the model may be more meaningful when enforcing a \emph{structural hierarchy} that only allows to estimate moderation effects to be nonzero if all respective pairwise interactions are estimated nonzero \citep{bien2013lasso}. However, such a choice reduces the fit of the model and should be based on substantive grounds.

Third, we use a single regularization parameter $\lambda_s$ in each of the $p$ nodewise regressions. The sparsity assumption here is with respect to the entire parameter vector, which includes both pairwise interactions and moderation effects (3-way interactions). However, if one assumes a different level of sparsity for pairwise interactions and moderation effects, separate regularization parameters may be more appropriate. The downside of two regularization parameters is that any model selection procedure needs to search a grid of $\lambda$s instead of a sequence, which increases the computational cost. In addition, estimates may become more unstable, because the considered model space is larger.

Fourth, it would be useful to extend the present work to moderated Mixed Graphical Models (MGMs) \citep{chen2014selection, yang2014mixed}. Here we have shown how to extend the multivariate Gaussian distribution by adding linear moderation effects for one or several moderator variables. This approach of adding terms for moderation effects to the nodewise regression equations can in principle be extended to the more general class of MGMs, in which each variable in the model is a univariate member of the exponential family conditioned on all other variables \citep{chen2014selection, yang2014mixed}. The only difference is that we perform the nodewise regressions in the GLM framework using the appropriate link-functions \citep[see e.g.][]{nelder1972generalized}. However, depending on which types of variables are involved in an interaction, both pairwise interactions and moderation effects are captured by \emph{sets} of varying numbers of parameters. It would be useful to give detailed treatment of all possible (moderated) interaction types between different types of variables, a description of how to interpret them and provide performance results for estimating different types of moderation effects in different situations. A special case of moderated MGMs would be an MGM with a single categorical variable as a moderator. This would allow to investigate group differences across two or \emph{several} groups in a principled way and is likely to out-perform group-split based methods on this task. Note that these analyses are implemented in the R-package \emph{mgm} \citep{haslbeck2017mgm} which we used in the application section of this paper. However, a full treatment of moderated MGMs is beyond the scope of the present paper.

Fifth, it could be useful to extend the notion of centrality to factor graphs and obtain a \emph{moderator centrality}. A naive way of doing that would be to simply add up the moderation effect of a given variable to find out which variable has the strongest influence on the pairwise interactions in the network model. But one could come up with more sophisticated measures that take the structure of the network into account. This would especially interesting in data sets that include contextual variables, because it allows to identify which of them have the strongest influence on a network model of psychological variables or symptoms.

In sum, Moderated Network Models relax the assumption of Gaussian Graphical Models that each pairwise interaction is independent of the value of all other variables by allowing that each pairwise interaction is \emph{moderated} by (potentially) all other variables. This allows more precise statements about the sign and value of a given interaction parameters in a given situation, which may reduce the presence of seemingly contradictory research outcomes and provides a step towards more accurate models for subgroups and individuals.

\section*{Acknowledgements}

We would like to thank George Aalbers, Fabian Dablander, Peter Edelsbrunner, Sacha Epskamp, Eiko Fried, Oisin Ryan for helpful comments on earlier versions of this paper.

\bibliographystyle{plainnat}
\bibliography{modNW_bib}

\appendix

\section{Mean-centering of Predictors in Moderation Analysis}\label{sec_linreg_centering}

In the procedure to estimate MNM described in Section \ref{methods_nodewisereg} we mean-center all variables before estimation. The reason is that for centered variables, the interpretation of parameters is more meaningful. We illustrate this issue in Figure \ref{fig_moderation_centering}:

\begin{figure}[H]
	\centering
	\includegraphics[width=1\textwidth]{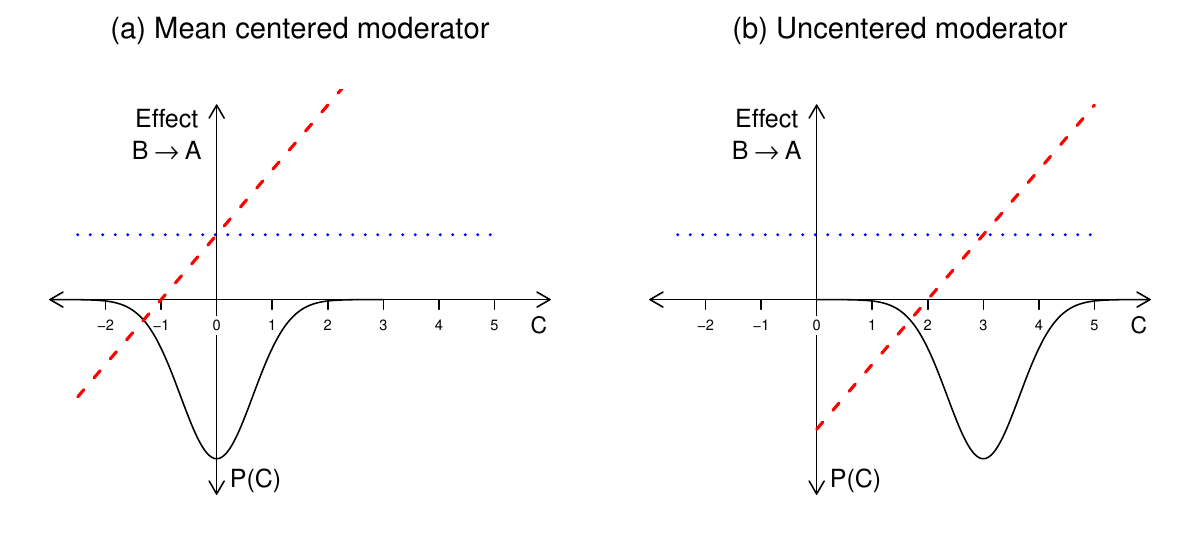}
	\caption{Illustration of the advantages of mean-centering predictors in moderated regression. For the mean-centered moderator in (a) the effect of $B$ on $A$ for $C=0$ is the same as the effect of $B$ on $A$ when ignoring $C$ (dotted blue line), which allows to compare parameters in models with/without moderation. In (b) this is not the case. Indeed, the effect of $B$ on $A$ if $C=0$ is an effect that is hardly ever observed because the probability of observing values of $C$ close to 0 is extremely small.
	}\label{fig_moderation_centering}
\end{figure}

In Figure \ref{fig_moderation_centering} (a), the red dashed line represents the partially moderated effect which we already considered above in Figure \ref{fig_moderation_simple} (d). In this case the moderator $C$ was mean-centered and its mean is therefore equal to zero. Recall the interpretation of the parameter $\beta_B$ in (\ref{eq_multiple_regression_interaction_2}): it is the effect of $B$ on $A$ when $C=0$. The blue dotted line represents the effect of $B$ on $A$ when averaging over all values of $C$, which is the effect of $B$ on $A$ one would obtain from a regression \emph{without} moderation effects. We see that the red and blue line intersect at $C=0$: this means that $\beta_{B}$ in the moderated regression in  (\ref{eq_multiple_regression_interaction_2}) has the same value as $\beta_{B}$ in the regression without moderation in (\ref{eq_multiple_regression}). This is desirable because it allows to compare the parameters in  models with/without moderation. Second, mean-centering moderators ensures that the parameter $\beta_{B}$ is meaningful in the sense that the probability of $C$-values close to zero is large. This is in contrast to Figure \ref{fig_moderation_centering} (b) in which the mean of the moderator $C$ is equal to 3. Now the effect of $B$ on $A$ when $C=0$ does not intersect with the blue line at the appropriate level. Therefore, $\beta_{B}$ captures the effect of $B$ on $A$ for values of $C$ close to zero which occur with an extremely low probability and are therefore irrelevant.

For a detailed discussion of interaction/moderation effects in linear regression we refer the reader to \cite{aiken1991multiple} and \cite{afshartous2011key}.

\section{Joint distribution for $p = 3$}\label{app_HOI_and_norm}

Here we show for the case of $p = 3$ variables how to factorize $p$ conditional distributions to a joint distribution.

We begin with the standard formulation of the conditional univariate Gaussian distribution

$$
P(X_1 | X_2 = x_2, X_3 = x_3) = \frac{1}{\sqrt{2\pi \sigma^2}}  
\exp \left \{  
- \frac{(X_1 - \mu_1)^2}{2\sigma^2}
\right \},
$$

\noindent
where the mean of $X_1$, $\mu_1$ is a function of $X_2$ and $X_3$. If we let $\sigma = 1$ and expand $(X_1 - \mu_1)^2$ we get

$$
P(X_1 | X_2 = x_2, X_3 = x_3) = \frac{1}{\sqrt{2\pi }}  
\exp \left \{  
- \frac{X_1^2 + \mu_1^2 - 2X_1\mu_1}{2}
\right \},
$$

\noindent
and can rearrange

$$
P(X_1 | X_2 = x_2, X_3 = x_3) = \frac{1}{\sqrt{2\pi }}  
\exp \left \{  
X_1\mu_1
- \frac{X_1^2}{2} 
- \frac{\mu_1^2}{2}
\right \}.
$$

Our focus is on $\mu_1$, so we absorb $ \frac{1}{\sqrt{2\pi }}$ and $- \frac{\mu_1^2}{2}$ in the log-normalizing constant $\Psi_1(\alpha, \beta, \omega)$ and let $C_1 = \frac{X_1^2}{2} $

\begin{equation}\label{eq_app_conditional}
P(X_1 | X_2 = x_2, X_3 = x_3) = 
\exp \left \{  
X_1\mu_1
- C_1
- \Psi_1(\alpha, \beta, \omega)
\right \}
\end{equation}

\noindent
where $\alpha, \beta,$ and $\omega$ are the parameter vectors defining the mean $\mu_1$, which is a linear combination of the other variables $X_2, X_3$:

\begin{equation}\label{eq_app_conditional_mean}
\mu_1 = \alpha_1 + \beta_{2,1} X_2 + \beta_{3,1} X_3 + \omega_{2,3,1} X_2 X_3
\end{equation}

\noindent
where $ \alpha_1$ is the intercept, $\beta_{2,1}, \beta_{3,1}$ are the parameters for the pairwise interactions with $X_2$ and $X_3$, and $ \omega_{2,3,1}$ is the parameter for the three-way interaction with $X_2 X_3$ (or equivalently, the pairwise interaction with $X_2$ moderated by $X_3$ or the pairwise interaction with $X_3$ moderated by $X_2$).

Plugging the mean (\ref{eq_app_conditional_mean}) into the conditional distribution (\ref{eq_app_conditional}) gives us

\begin{align*}
	P(X_1 | X_2 = x_2, X_3 = x_3) = 
	\exp \left \{  
	X_1 (\alpha_1 + \beta_{2,1} X_2 + \beta_{3,1} X_3
 \right. 
\\ \nonumber
\left. 
+ \omega_{2,3,1} X_2 X_3)
	- C_1
	- \Psi_1(\alpha, \beta, \omega)
	\right \}.
\end{align*}

Multiplying out and collecting terms of the same order gives

\begin{align*}
	P(X_1 | X_2 = x_2, X_3 = x_3) = 
\exp \{  
\alpha_1 X_1 +
X_1 \sum_{\substack{j=1, j\neq 1}}^3 \beta_{i, j}  X_j	  
\\ 
+ X_1 \omega_{1,2,3} X_2 X_3
- C_1
- \Psi_1(\alpha, \beta, \omega)
\}
.
\end{align*}

Similarly define $P(X_2 | X_1 = x_1, X_3 = x_3) $ and $P(X_3 | X_2 = x_2, X_1 = x_1)$. Then factorize the three conditional distributions to obtain a joint distribution. After rearranging terms we get

\begin{align}\label{eq_app_joint}
P(X_1, X_2, X_3) = 
\exp \left \{  
\sum_{i = 1}^{3} \alpha_1 X_i +
\sum_{i = 1}^{3}\sum_{\substack{j=1, j\neq i}}^3 \beta_{i, j} X_i X_j	  
 \right. 
\\ \nonumber
\left. 
+ \omega_{1,2,3} X_1 X_2 X_3
- 
\sum_{i = 1}^{3} [C_i + \Psi_i(\alpha, \beta, \omega)]
\right \},
\end{align}

\noindent
where we combined the parameters $\beta_{i,j}, \beta_{j,i}$ and $\omega_{j,i,z}, \omega_{i,j,z}, \omega_{z,j,i}$ into single parameters by taking their average.

The joint distribution can be constructed analogously for any $p$.

A sufficient condition for (\ref{eq_app_joint}) to be normalizable is that the sum over all terms in the exponential is negative \citep{yang2014mixed}. However, the constraints $\mathcal{C}(\alpha, \beta, \omega)$ on the parameter space $\{\alpha, \beta, \omega\}$ to ensure this, are unknown. Note that these constraints are possibly very complicated since they depend on the variances of the conditional distributions and the structure of the Factor graph defining this higher order model \citep{koller2009probabilistic}. To be able to sample from (\ref{eq_app_joint}), we use a rejection sampler as described in Appendix \ref{app_Sampling}.

\section{Rejection Sampling}\label{app_Sampling}

We do not know the constraints $\mathcal{C}(\alpha, \beta, \omega)$ on the parameter space that ensure that the model in (\ref{eq_app_joint}) (and it's generalization to $p$ variables) is normalizable. To still be able to sample from this distribution we use a rejection sampler that rejects diverging chains in the Gibbs sampler \cite{casella1992explaining}. Specifically, we fix the MNM and sample cases using the Gibbs sampler, with a burn-in of 100 iterations. A chain is defined to diverge if $| X_i | > \tau$ for at least one $i \in \{1, 2,  \dots, p\}$, where we set $\tau = 3.09$, which is the $99.9$\% quantile of a standard normal distribution.

This way we define $\mathcal{C}(\alpha, \beta, \omega)$ indirectly. If the chain remains within $[-\tau, \tau]$ we assume that  the constraints $\mathcal{C}(\alpha, \beta, \omega)$ are satisfied. If the chain diverges, we know that $\mathcal{C}(\alpha, \beta, \omega)$ is not satisfied. Note that diverging chains approach $\pm \infty$ very quickly, therefore the exact value of $\tau$ has only a small impact on the sampling procedure.

With this procedure it would be possible to sample cases for any fixed MNM generated with the procedure described in Section \ref{sec_simulation_datageneration}. However, for some MNM the proportion of rejected samples might be high, and consequently the running time until obtaining $n=1808$ cases would be very large. To keep the computational manageable, we sampled  $n=10000$ cases from 130 models as specified in Section \ref{sec_simulation_datageneration}. We then ordered the 130 population models by increasing proportion of diverged samples, and selected the first 100. In those 100 iterations the proportion of rejected samples varied between $0.287$ and $0.886$. We take the $n=1808$ first observations in each of the 100 data sets. These data are used in the simulation study.

To check whether the rejection sampling introduced bias in the estimates, we recover every single parameter in the 100 data sets,  using standard linear regressions in which we specified the correct moderator. Within data sets we take the average over pairs of parameter types (unmoderated, partially moderated, fully moderated) estimated from $n=1808$ observations, and plot them as a function of the proportion of finite samples in the given iteration in Figure \ref{fig_app_sanitycheck}:

\begin{figure}[H]
	\centering
	\includegraphics[width=.9\textwidth]{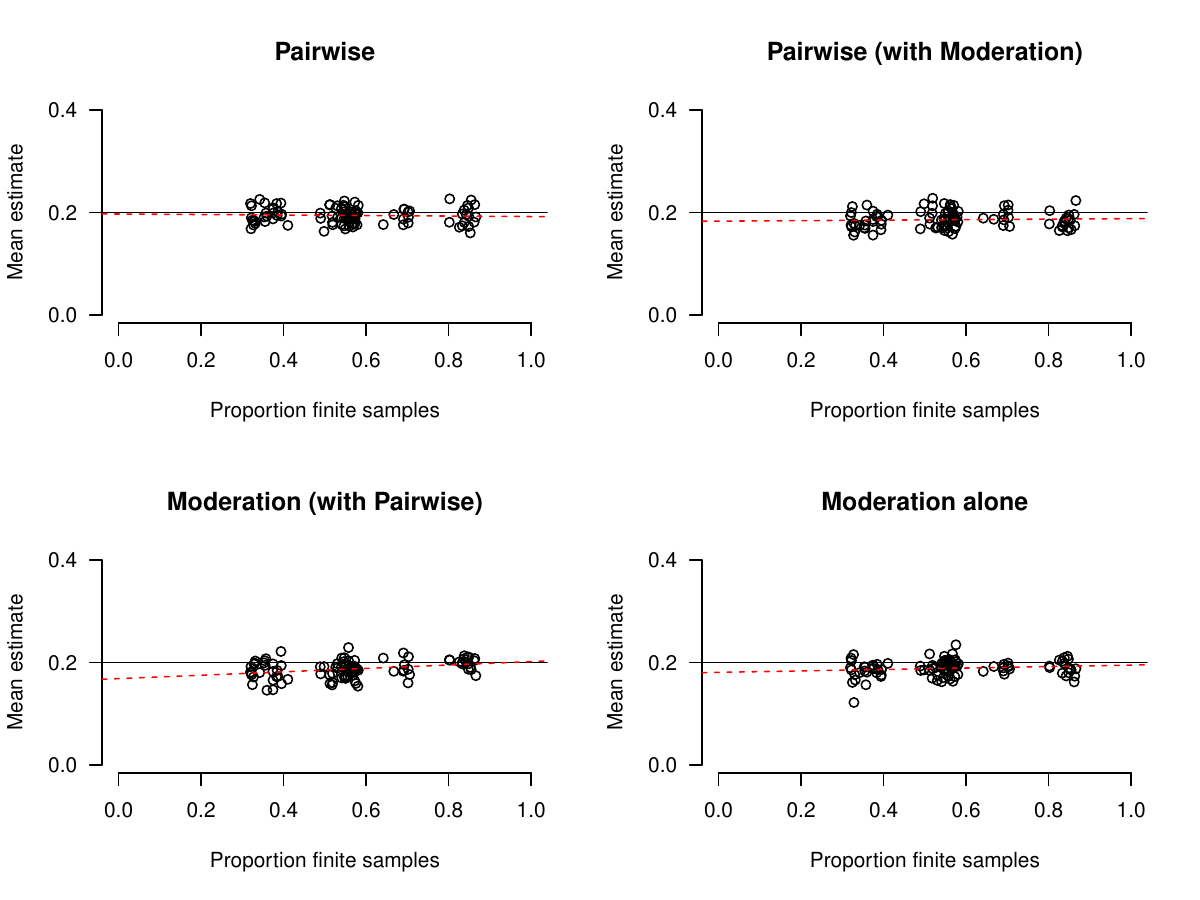}
	\caption{Each point is the average over the two estimates of a given type in a given iteration, estimated with multiple regression with the correct moderator specified.}\label{fig_app_sanitycheck}
\end{figure}

Figure \ref{fig_app_sanitycheck} shows that the unmoderated pairwise interactions are slightly biased downwards ($\bar{\beta}_{pw} = 0.194$). The remaining three parameter types show a stronger downward bias ($\bar{\beta}_{pw(mod)} = 0.186$, $\bar{\omega}_{mod(pw)} = 0.187$ and $\bar{\omega}_{mod} = 0.189$). The bias of the moderation effect seems to decrease with increasing proportion of finite samples. However, all parameter estimates are close to the value specified in the Gibbs sampler and hence the data can be used in the simulation. The result of this slight downward bias is that the performance across all conditions and estimators is also slightly biased downward.

\section{Comparing Sensitivity of Moderated Network Model across Parameter Types}\label{app_results_compareTypes}

In Section \ref{sec_sim_results} we claimed that the sensitivity to detect unmoderated pairwise interactions (row 1 in Figure \ref{fig_sim_results_part1}) and full moderations (row 2 in Figure \ref{fig_sim_results_part2}) was lower than for detecting the pairwise interaction (row 2 in Figure \ref{fig_sim_results_part1}) and moderation effect (row 2 Figure \ref{fig_sim_results_part2}) in the partially moderated pairwise interaction. This is not easy to see in these figures and we therefore provide an additional figure with this comparison in Figure \ref{fig_app_SENacrossT}:

\begin{figure}[H]
	\centering
	\includegraphics[width=1\textwidth]{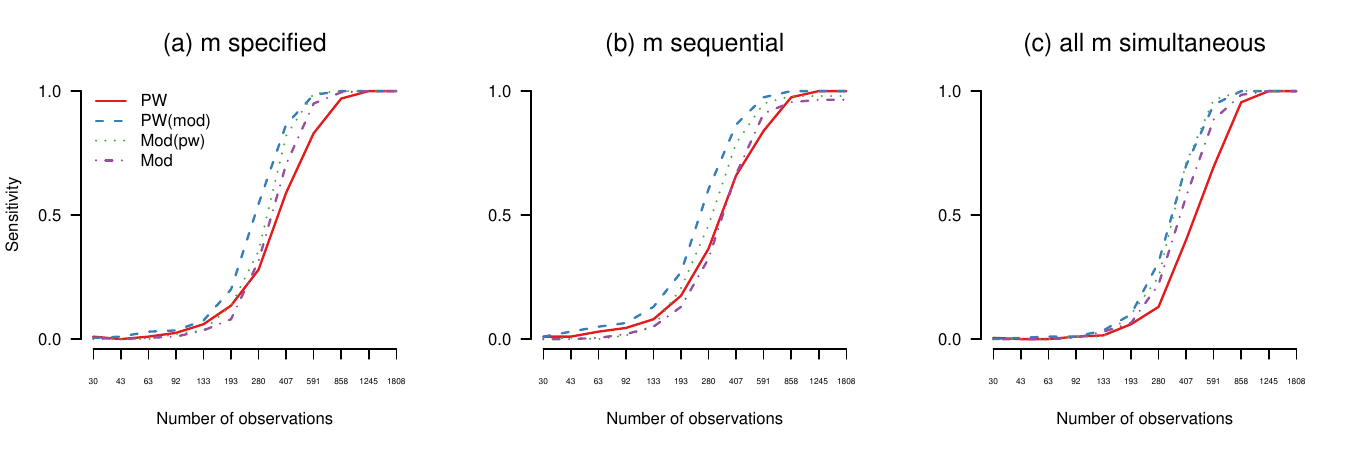}
	\caption{Sensitivity to detect the four different parameter types, separately for the moderated network model with (a) correctly specified moderator, (b) sequentially searching all moderators and (c) specifying all moderators at once.}\label{fig_app_SENacrossT}
\end{figure}

To explain this difference, we first run a simpler simulation in which we isolate all interaction types to exclude the possibility that the differences are explained by some characteristic of the graph (Appendix \ref{app_basicSim}). We will find that in such a "clean" setting the difference is even larger. Then we explain this difference in terms of the number of uncorrelated neighbors (Appendix \ref{app_results_neighbors}).

\section{Simulation with Isolated Interaction Types}\label{app_basicSim}

Here we run a simplified version of the simulation reported in Section \ref{sec_simulation} to exclude the possibility that the sensitivity differences discussed in Appendix \ref{app_results_compareTypes} can be explained by some graph characteristic. We generate data from the graph shown in Figure \ref{fig_app_basicSim_setup} panel (a):

\begin{figure}[H]
	\centering
	\includegraphics[width=1\textwidth]{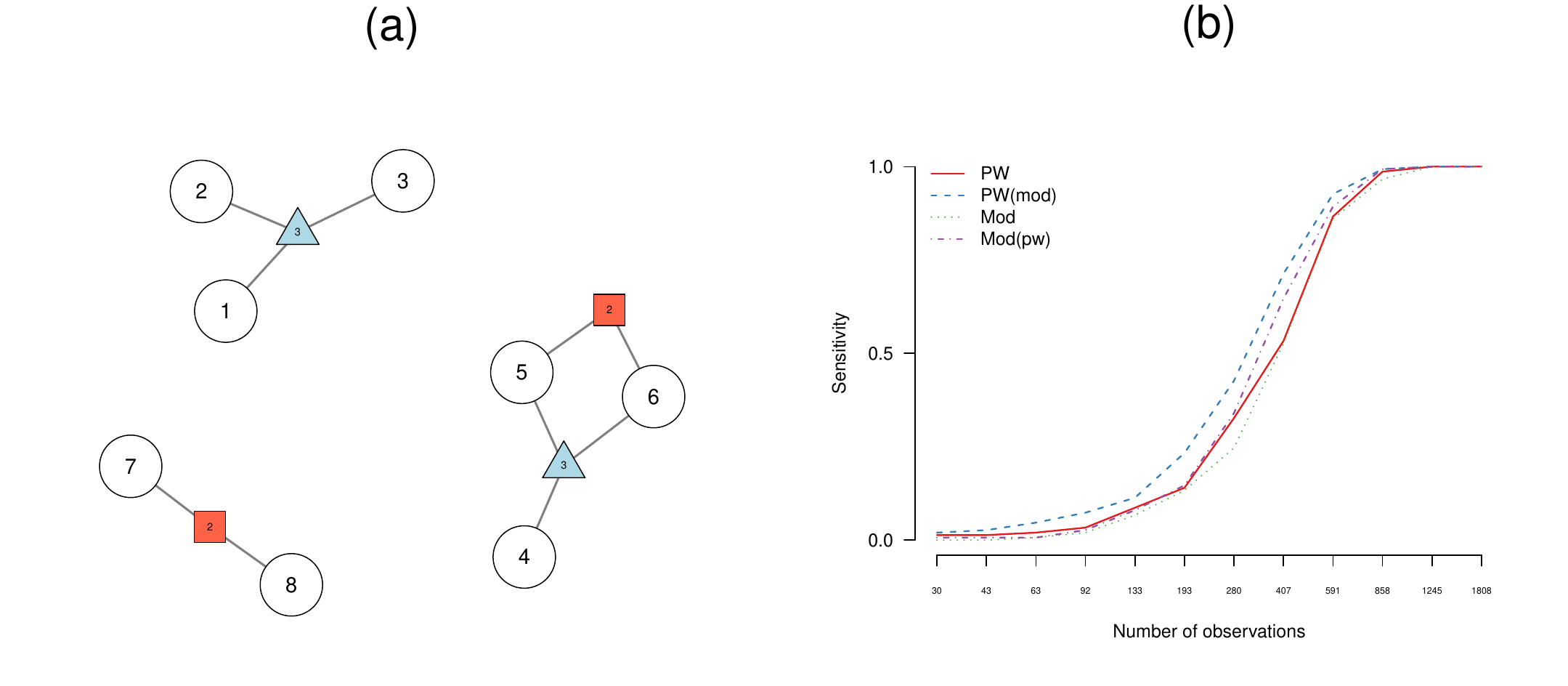}
	\caption{Panel (a): simplest possible graph including all four parameter types that separates all types as far as possible; panel (b): sensitivity to recover each parameter type for different numbers of observation for the moderated network model with specified moderator.}\label{fig_app_basicSim_setup}
\end{figure}

Except the graph structure, we use the same setup as reported in Section \ref{sec_simulation_datageneration}. In this graph the four parameter types are isolated as much as possible, so that the graph structure does not have any influence on estimation:  7-8 is isolated; the fully moderated pairwise interaction (or 3-way interaction) 1-2-3 is isolated; and the partially moderated pairwise interaction 4-5-6 is isolated. Note that the pairwise interaction 5-6 and the moderation effect 4-5-6 cannot be separated by definition.

We estimated the moderated network model with specified moderator (here variables 4 and 1). If we find the same sensitivity difference in this simulation as in Appendix \ref{app_results_compareTypes}, we know that it is not a function of some unexpected graph characteristic. Figure \ref{fig_app_basicSim_setup} panel (b) shows the sensitivity for the four parameter types as a function of the number of observations. We observe the same sensitivity difference as in Appendix \ref{app_results_compareTypes}.

What is the difference between the pairwise interaction 7-8 and the moderation effect (or 3-way interaction) 1-2-3 on the one hand, and the pairwise interaction and the moderation effect in 4-5-6 on the other hand? The difference is that in the respective nodewise regressions, in the former case there is one nonzero predictor and in the latter case there are two nonzero predictors (see equation (\ref{eq_multiple_regression_interaction_1}) in Section \ref{sec_methods_simplemoderation}). The presence of two nonzero predictors results in that the EBIC selects a smaller (compared to the presence of one nonzero predictor) penalty parameter $\lambda_s$, which increases sensitivity. This reasoning is based on the assumption that the predictors are uncorrelated, which is the case if all variables and interaction effects are centered as in the present case. In Appendix \ref{app_centering}, we show that $X$ is uncorrelated with $XY$ if both $X$ and $Y$ are centered. In Appendix \ref{app_results_neighbors}, we test this explanation directly by investigating sensitivity as a function of the number of uncorrelated neighbors in a GGM.

\section{Sensitivity as Function of Number of Uncorrelated Neighbors}\label{app_results_neighbors}

Here the goal is to directly test the hypothesis that the sensitivity to detect the edges between nodes $X_1$ and $X_2, \dots, X_p$ increases with the number of nonzero edges, if $X_2, \dots, X_p$ are uncorrelated. To this end we generate a GGM with $p=20$, which matches the maximum neighborhood size of each node to that of the simplified simulation in Appendix \ref{app_basicSim}. We compare four versions of this GGM, across which we vary the number of uncorrelated neighbors from 1-4. The 1-4 nonzero partial correlations have all the value $\beta = 0.2$, matching the setup of the simulations in Section \ref{sec_simulation} and Appendix \ref{app_basicSim}. We show the average (over 1-4 edges in the four conditions, respectively) sensitivity in Figure \ref{fig_app_neighbors_results}:

\begin{figure}[H]
	\centering
	\includegraphics[width=.65\textwidth]{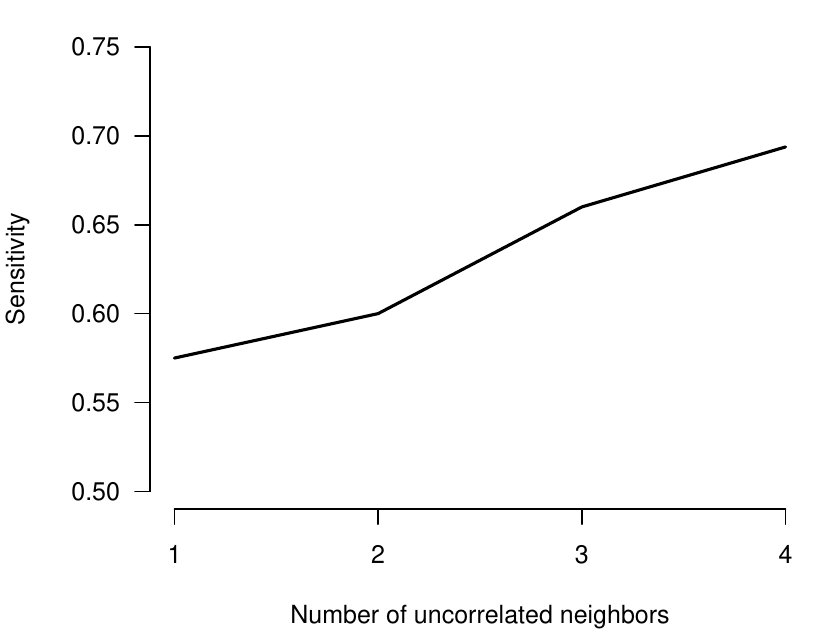}
	\caption{Sensitivity to detect a neighbor connected to node $X_1$ as a function of the number of uncorrelated neighbors of $X_1$. }\label{fig_app_neighbors_results}
\end{figure}

The results confirm the hypothesis that the sensitivity to detect neighbors increases as a function of the number of uncorrelated neighbors.

\section{XY is uncorrelated with X and Y if the latter are mean-centered}\label{app_centering}

For independent and identically distributed variables $X, Y$ with finite variances, let $Z = XY$. We show $\rho(X, Z) = \frac{\text{cov}\{X, Z\} } {\sigma_{X} \sigma_{Z}}  = 0$. Since $\sigma_{X}, \sigma_{Z} > 0$, $\rho(X, Z) = 0$ iff $\text{cov}(X, Z) = 0$. However,

\begin{align*}
	\text{cov}(X, Z) &= \mathbb{E}[XZ] - \mathbb{E}[X] \mathbb{E}[Z] \\
	&= \mathbb{E}[X^2Y] - \mathbb{E}[X] \mathbb{E}[XY] \\
	&= \mathbb{E}[X^2] \mathbb{E}[Y] - \mathbb{E}[X] \mathbb{E}[X]  \mathbb{E}[Y] \\
	&= (\mathbb{E}[X^2] - (\mathbb{E}[X])^2) \mathbb{E}[Y] \\
	&= \sigma_{X}^2 \mathbb{E}[Y],
\end{align*}

\noindent
so $\rho(X, Z)=0$ if $\mathbb{E}[Y] = 0$, and similarly, $\rho(Y, Z)=0$ if $\mathbb{E}[X] = 0$.

\section{Additional Tutorial: Estimate MNM on Iteration 2 of Simulation Study}\label{app_simdata_tutorial}

In this section, we show how to use the R-package \emph{mgm} \citep{haslbeck2017mgm} to fit a moderated network model to $n = 858$ observations generated from the model shown in Figure \ref{fig_sim_generating}. The \emph{mgm} implements functions to estimate Mixed Graphical Models (MGMs), of which GGMs are a special case. The package can be installed and loaded in the following way:

\begin{verbatim}
install.packages("mgm")
library(mgm)
\end{verbatim}

We will also present different options for visualizing the moderated network model using factor graphs.

\subsection{Fit Moderated Network Model to Data}

The data set is automatically available in the list object \verb|modnw| when loading the \emph{mgm} package. As specified in Section \ref{sec_simulation_datageneration}, the data set contains 13 continuous variables and we sampled 858 observations:

\begin{verbatim}
> dim(modnw)
[1] 858  13
\end{verbatim}

We provide the data in \verb|modnw| to the estimation function \verb|mgm()| of the \emph{mgm} package.  Next to the data we specify the types and levels for each variable. Since we model all variables as continuous Gaussian distributions, we specify \verb|"g"| for each variable and the number of levels as \verb|1| by convention for continuous variables. This specification is necessary in \emph{mgm}, because the package also allows to model Poisson variables and categorical variables with $k$ categories. Via the argument \verb|moderator| one specifies the moderators to be included in the model. For instance, if we select \verb|moderator = c(3, 7)| all moderation effects of variables 3 and 7 are included in the model. Here we pretend not to know that variable 13 is the only moderator in the model and and therefore include all variables as moderators by setting \verb|moderator = 1:13|. 

The estimation algorithm uses $p$ nodewise penalized regressions, for each of which an appropriate regularization parameter $\lambda_s$ has to be selected (see Section \ref{methods_nodewisereg}). We select the $\lambda_s$ that minimizes the EBIC with the hyperparameter $\gamma = 0.5$ by setting \verb|lambdaSel = "EBIC"| and  \verb|lambdaGam = .5|. Alternatively one could select $\lambda_s$ using cross-validation (\verb|lambdaSel = "CV"|). With \verb|scale = TRUE| we specify that all predictors are scaled to mean zero and SD = 1. This is a standard procedure in regularized regression and avoids that the penalization of a given parameter depends on the standard deviation of the associated variable. With \verb|ruleReg = "AND"| we specify that the nodewise regressions are combined with the AND-rule (see Section \ref{methods_nodewisereg}).

\begin{verbatim}
mgm_mod <- mgm(data = modnw, 
               type = rep("g", 13),
               level = rep(1, 13),
               moderator = 1:13,
               lambdaSel = "EBIC", 
               lambdaGam = .5,
               ruleReg = "AND",
               scale = TRUE)
\end{verbatim}

The main output is stored in \verb|mgm_mod$interactions|. For a detailed description of the output see the help file \verb|?mgm| and the \emph{mgm} paper \citep{haslbeck2017mgm}. The list entry \verb|mgm_mod$interactions$indicator| contains a list of all estimated parameters separately for each order (2-way, 3-way, etc.):

\begin{verbatim}
> mgm_mod$interactions$indicator
[[1]]
[,1] [,2]
[1,]    1   12
[2,]    2    4
[3,]    4   11
[4,]    8   11

[[2]]
[,1] [,2] [,3]
[1,]    1   12   13
[2,]    6    7   13
[3,]    8   10   13
[4,]    8   11   13
\end{verbatim}

The first level contains pairwise (2-way) interactions and the second entry contains moderation effects (or 3-way interactions). Thus, in the above output the entry \verb|mgm_mod$interactions$indicator[[1]][2, ]| indicates that there is a nonzero pairwise interaction between variables 2-4. And the entry \verb|mgm_mod$interactions$indicator[[2]][4, ]| indicates that there is a nonzero moderation effect (or 3-way interaction) between variables 8-11-13. In the present model we estimated four pairwise interactions and four moderation effects. To obtain more information about a given interaction we use the function \verb|showInteraction()|. Here is how to obtain the parameter for the pairwise interaction 4-11:

\begin{verbatim}
> showInteraction(object = mgm_mod, int = c(4,11))
Interaction: 4-11 
Weight:  0.103782 
Sign:  1 (Positive)
\end{verbatim}

The pairwise interaction can be interpreted as in a linear regression: when increasing $X_4$ by one unit, $X_{11}$ increases by $\approx 0.104$ units, when keeping all other variables constant. The parameters for the moderation effects can be obtained similarly: let's say we are interested in the moderation effect 6-7-13. Then we obtain the absolute value and the sign of the parameter via:

\begin{verbatim}
> showInteraction(object = mgm_mod, int = c(6,7,13))
Interaction: 6-7-13 
Weight:  0.1174669 
Sign:  1 (Positive)
\end{verbatim}

We can interpret this moderation effect in the following way: the pairwise interaction between $X_6$ and $X_7$ is zero when $X_{13}$ is equal to zero. When increasing $X_{13}$ by one unit, the pairwise interaction between $X_6$ and $X_7$ is equal to $\approx 0.117$. Similarly, this parameter can be interpreted the moderation effect of $X_6$ on the pairwise interaction between $X_7$  and $X_{13}$, or the moderation effect of $X_7$ on the pairwise interaction between $X_6$ and $X_{13}$.

The interpretation is slightly different if a variable is involved in a partially moderated pairwise interaction (or equivalently, in both a 2-way and 3-way interaction). We take variable 12 as an example. We have a pairwise interaction

\begin{verbatim}
> showInteraction(object = mgm_mod, int = c(1,12))
Interaction: 1-12 
Weight:  0.1269736 
Sign:  1 (Positive)
\end{verbatim}

\noindent
and a moderation effect:

\begin{verbatim}
> showInteraction(object = mgm_mod, int = c(1,12,13))
Interaction: 1-12-13 
Weight:  0.1467114 
Sign:  1 (Positive)
\end{verbatim}

In this case the pairwise interaction between $X_1$ and $X_{12}$ is equal to $\beta_{1, 12} \approx 0.130$ if $X_{13}=0$. If $X_{13}$ increases one unit, then the pairwise interaction between $X_1$ and $X_{12}$ increases by $\approx 0.147$, so $\approx 0.130 + 1 \cdot 0.147$.

\subsection{Moderated Network Model as Factor Graph}

Moderation effects (3-way interactions) cannot be visualized in a standard graph. However, they can be visualized in a Factor graph, which introduces a new node for each interaction parameter. Such a Factor graph can be drawn using the \verb|FactorGraph()| function that takes the output of  \verb|mgm()| as input:

\begin{verbatim}
FactorGraph(object = mgm_mod,
PairwiseAsEdge = FALSE)
\end{verbatim}

The \verb|FactorGraph()| function plots the graph visualization in Figure \ref{fig_appl_factorgraph_app} (a):

\begin{figure}[H]
	\centering
	\includegraphics[width=1\textwidth]{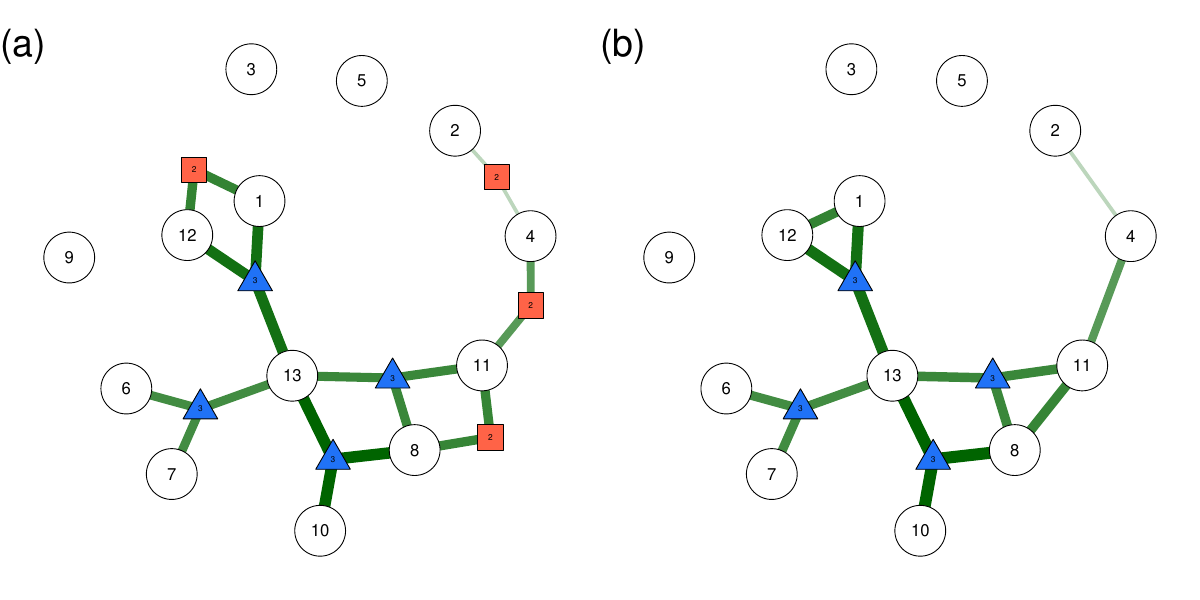}
	\caption{Two different Factor graph visualizations: (a) variable-nodes are displayed as circle nodes, pairwise interactions are displayed as square nodes, and moderation effects (3-way interactions) are displayed as triangles; (b) Only moderation effects (3-way interactions) are displayed as triangle nodes, pairwise interactions are displayed as simple edges. Green edges indicate parameters with positive sign. The widths of edges is proportional to the absolute value of the parameter. }\label{fig_appl_factorgraph_app}
\end{figure}

The green (red) edges indicate parameters with positive (negative) sign, and the width of edges is proportional to the absolute value of the parameter. \verb|FactorGraph()| is a wrapper around the \verb|qgraph()| function from the \emph{qgraph} package \citep{epskamp2012qgraph} and all \verb|qgraph()| arguments can passed to customize the visualization.

In larger graphs with many pairwise interactions this visualization may become unclear. For these situations, the factors representing pairwise interactions can be replaced by simple edges by setting \verb|PairwiseAsEdge = TRUE|. The resulting visualization is shown in Figure \ref{fig_appl_factorgraph_app} (b). While this graph is not a typical Factor graph anymore, the visualization contains the same information as the visualization in (a).

\section{Varying Moderation Effects across Nodewise Regresssions: A closer Look}\label{app_misspec_modeffects}

In Section \ref{sec_misspec_signs} we used a data set of three mood variables to illustrate that moderation effects can differ across nodewise regressions, if the data are skewed. Here we provide some intuition for how this is possible, by conditioning on different values of one of the three variables, and show the resulting conditional scatter plots and linear relationships of the remaining two variables.

We first estimate the three conditional distributions of the moderated network model using unregularized linear regression:

\begin{verbatim}
> lm(afraid~ashamed*distressed, data = msq_p3) # On afraid
Call:
lm(formula = afraid ~ ashamed * distressed, data = msq_p3)
Coefficients:
(Intercept)             ashamed          distressed  
-0.0481              0.1194              0.3210  
ashamed:distressed  
0.1142  

> lm(ashamed ~ afraid*distressed, data = msq_p3) # On Ashamed
Call:
lm(formula = ashamed ~ afraid * distressed, data = msq_p3)
Coefficients:
(Intercept)             afraid         distressed  
-0.04104            0.16279            0.25054  
afraid:distressed  
0.08581  

> lm(distressed ~ afraid*ashamed, data = msq_p3) # On Distressed
Call:
lm(formula = distressed ~ afraid * ashamed, data = msq_p3)
Coefficients:
(Intercept)          afraid         ashamed  afraid:ashamed  
0.01339         0.40273         0.30110        -0.02984  
\end{verbatim}

We estimated moderation effects with positive sign in the regressions on \emph{afraid} and \emph{ashamed} and a moderation effect with negative sign in the regression on \emph{distressed}. This is reflecting the results obtained from $\ell_1$-regularized regression shown in Figure \ref{fig_misspec_msq_p3_factorgraphs} in Section \ref{sec_misspec_signs}.

We visualize the phenomena of different moderation effects across nodewise regressions in Figure \ref{fig_misspec_app_1a} by conditioning on different values of one of the predictors and inspect the linear relationship between the response variable and the remaining predictor. We do this for each of the three regressions:

\begin{figure}[H]
	\centering
	\includegraphics[width=1\textwidth]{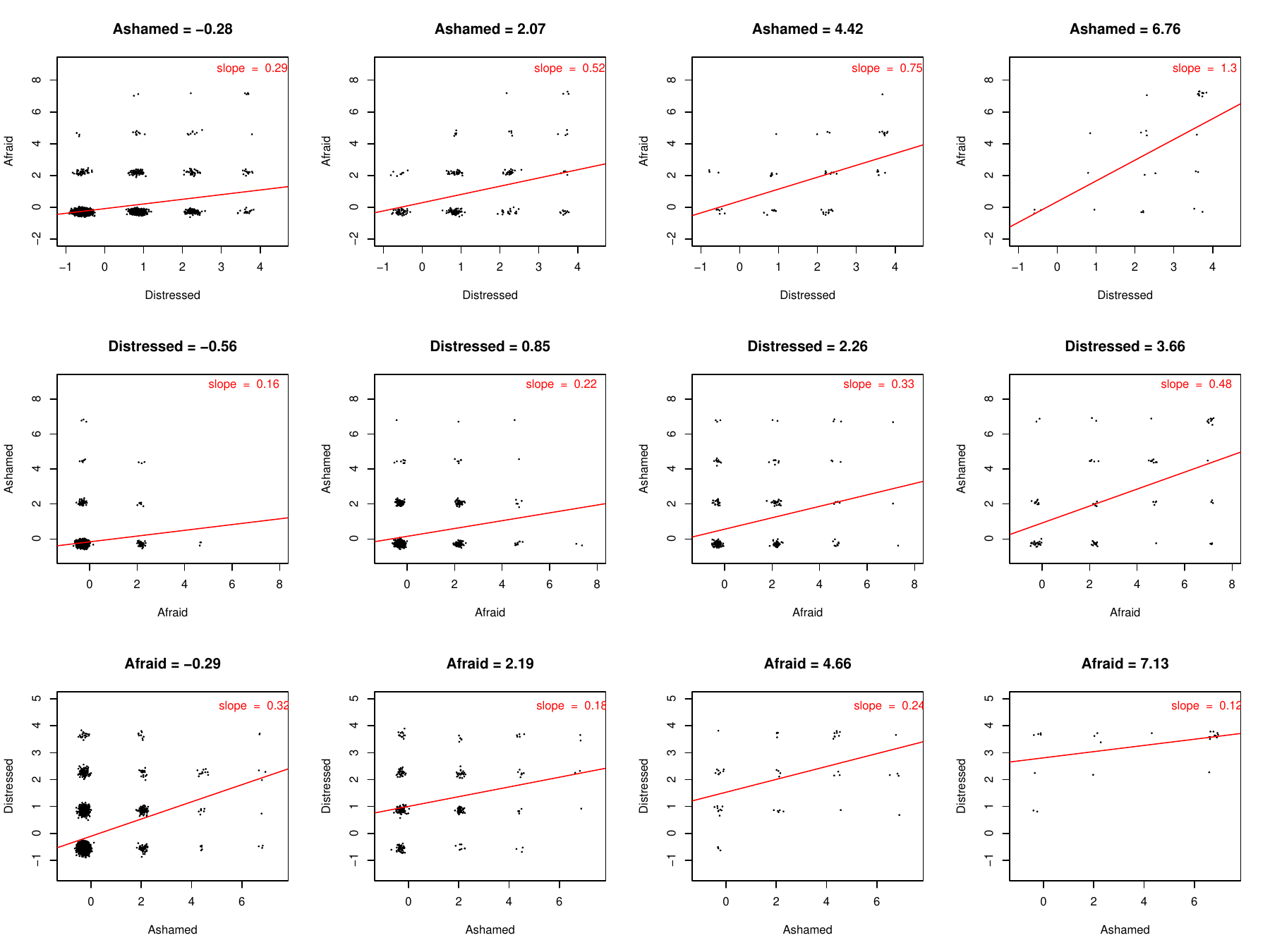}
	\caption{Row 1: The linear relationship between \emph{afraid} and \emph{distressed} depicted by a scatter plot and the best linear fit, for the four different values of \emph{ashamed}. We added some noise in the visualization to capture the amount of data at each combination of \emph{afraid} and \emph{distressed}; row 2: the same visualization as in row 1, however, for the regression on \emph{ashamed}, conditioning on the values of \emph{distressed}; and row 3: the same visualization as in row 1, however, for the regression on \emph{distressed}, conditioning on values of \emph{afraid}.}\label{fig_misspec_app_1a}
\end{figure}

The first row of Figure \ref{fig_misspec_app_1a} shows the scatter plot of variables \emph{afraid} and \emph{distressed} together with the best fitting regression line (red line) and its slope, for the different values of \emph{ashamed}. To make the density of the data visible we added a small amount of noise to each data point for the visualization. The best regression line was calculated on the original data. We see that the positive linear relationship between \emph{afraid} and \emph{distressed} becomes stronger for larger values of \emph{ashamed}. Similarly in row 2, the positive linear relationship between \emph{ashamed} and \emph{afraid} becomes stronger for larger values of \emph{distressed}. The linear relationship between \emph{afraid} and \emph{distressed} increases more as a function of \emph{ashamed}, than does the linear relationship between  \emph{ashamed} and \emph{afraid} as a function of \emph{distressed}. This reflects the fact that the moderation/interaction parameter in the regression on \emph{afraid} is larger than in the regression on \emph{ashamed}.

Finally, row 3 shows that the linear relationship between \emph{ashamed} and \emph{distressed} first decreases, then increases, and then decreases again as a function of \emph{afraid}. Thus, we see a non-linear moderation effect of \emph{afraid} on the linear relationship between \emph{ashamed} and \emph{distressed}. It happens to be the case that this non-linear moderation effect is not canceled out exactly, but is best approximated by a small negative linear moderation effect.

\section{Correlations between lower- and higher-order terms}\label{sec_misspec_colinearity}

If predictors $X$ and $Y$ are centered and independent distributions of any kind, we have $\text{cor} (X, XY) = 0$. That is, the lower order terms (singleton predictors such as $X$) and higher order terms (product term such as $XY$) are uncorrelated. We prove this in Appendix \ref{app_centering}. Also, if $X, Y$ are Gaussians and linearly dependent, $\text{cor} (X, XY) = 0$. However, if $X, Y$ are skewed \emph{and} linearly dependent, we have $\text{cor} (X, XY) \neq 0$. We illustrate these four cases (distribution symmetric vs. non-symmetric and dependent vs. independent $X$ and $Y$) in Figure \ref{fig_correlatedpredictors}:

\begin{figure}[H]
	\centering
	\includegraphics[width=.9\textwidth]{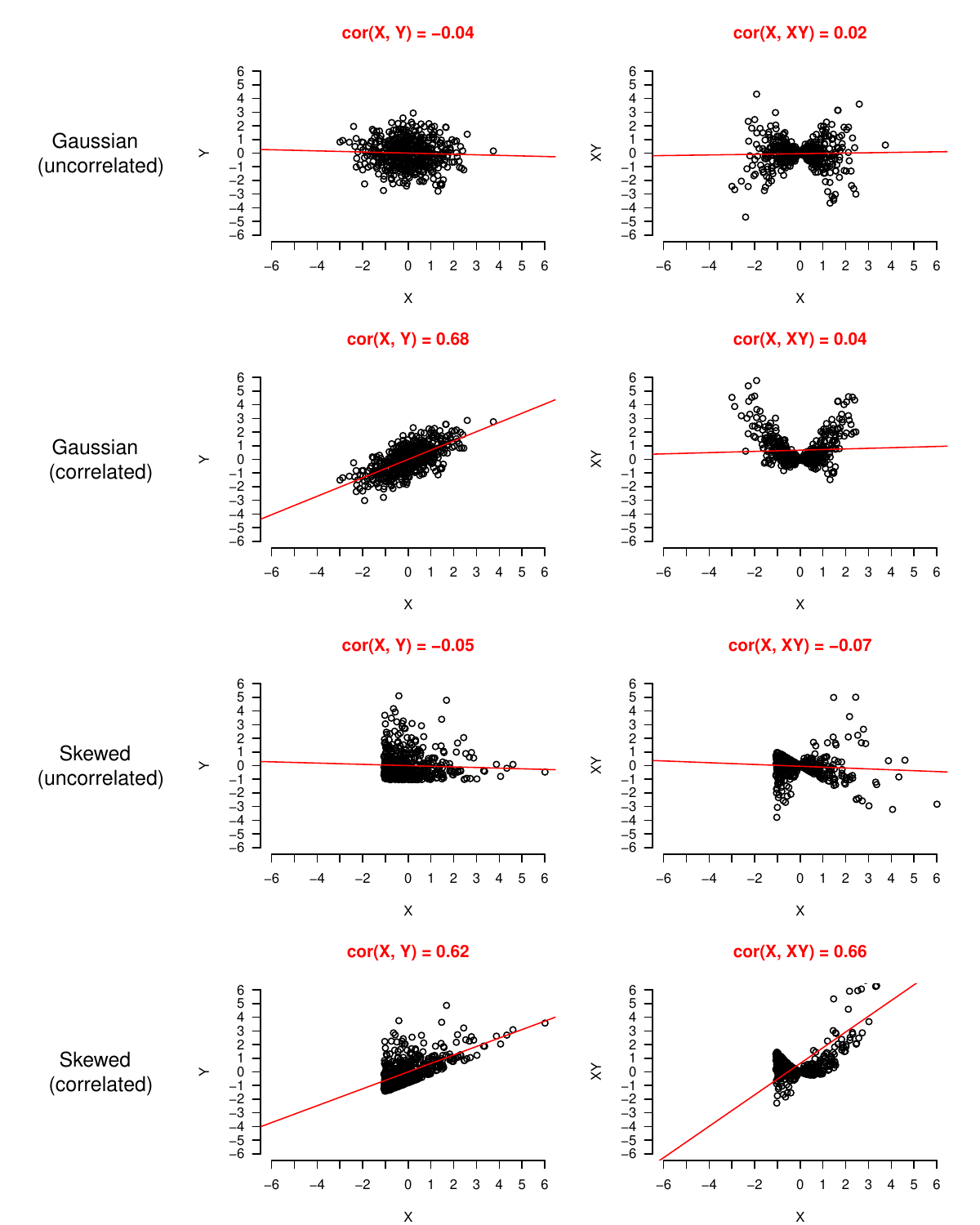}
	\caption{The correlation between lower- and higher-order terms $\text{cor}(X, XY)$ for the four combinations of symmetric (Gaussian) / skewed (Exponential) and correlated/uncorrelated (see text for details).}\label{fig_correlatedpredictors}
\end{figure}

In the first row we generated $n=500$ observations from two uncorrelated Gaussian distributions $X = \mathcal{N}(\mu = 0, \sigma = 1)$ and $Y = \mathcal{N}(\mu = 0, \sigma = 1)$. We then mean-centered both distributions. The first column shows the scatter plot, the best fitting regression line (red line) and the correlation $\text{cor}(X, Y)$. As we would expect the correlation is close to zero. We then plot $X$ against $XY$. And we see that the correlation $\text{cor}(X, XY)$ remains close to zero as claimed. In the second row we sampled from two correlated Gaussian distributions $X = \mathcal{N}(\mu = 0, \sigma = 1)$ and $Y = X + \mathcal{N}(\mu = 0, \sigma = 1)$ and mean-centered both distributions. As expected, we find a linear relationship between $X$ and $Y$. However, the correlation $\text{cor}(X, XY)$ remains close to zero. In the third row we sampled from two uncorrelated exponential distributions with rate parameter $\lambda = 1$, and mean-centered both distributions. We see that $\text{cor}(X, Y)$ is close to zero as expected. And also $\text{cor}(X, XY)$ is zero, as expected. Row four shows the problematic case. We correlated the exponential distributions by adding $X$ to $Y$, and mean-centered both distributions. Now we get a linear relationship between $X$ and $Y$ as expected. However, in contrast to the symmetric Gaussian case, this correlation now also implies a nonzero correlation $\text{cor}(X, XY)$ between $X$ and $XY$.

Uncorrelated predictors are a desirable property for two reasons: if correlations between predictors (also called collinearity) are very high, the model becomes unstable in the sense that changing only a few data points can substantially change the parameter estimates. Second, $\ell_1$-regularized regression has the property to select only one of several highly correlated predictors. If $\text{cor} (X, XY)$ is large the $\ell_1$-regularized estimator is likely to estimate a nonzero value for one of the two corresponding parameters, even though both parameters are nonzero in the true model. This problem is more severe for large regularization parameters $\lambda$ (which imply strong regularization) which tend to be selected if the number of observations $n$ is small.

The problem of correlated predictors can be diagnosed by correlating all lower-order terms with all higher-order terms. If they are highly correlated one solution is to transform the skewed variables towards a symmetric distribution, which reduces the correlation. Popular transformations are taking the log, the square-root or computing the nonparanormal transform \citep{liu2009nonparanormal}. However, transforming variables always has the disadvantage that a meaningful interpretation of relationships with transformed variables becomes difficult. For example, many people will have little intuition for relationships like "increasing \emph{Feelings of Guilt} by one unit increases log \emph{Mood} by 0.15 units". A solution for the problem of $\ell_1$-regularized (LASSO) estimators is to use an estimator that does not use an $\ell_1$-penalty. This could for instance be a regularized estimator with an  $\ell_2$-penalty (Ridge regression). But then we loose the convenient LASSO property of setting small parameters to zero with the result that every network (or factor graph) will be fully connected. Another option would be to use significance-test based estimators. However, then one would have to deal with the problem of multiple testing, which is a serious issues in moderated network models due to the potentially large number of estimated parameters. A detailed investigation of solutions to this particular problem of model misspecifications is beyond the scope though and we leave it for future research.

\section{Simulation Results in Tables}\label{app_results_in_tables}

In Section \ref{sec_simulation} we reported simulation results in Figures \ref{fig_sim_results_part1} and \ref{fig_sim_results_part2}. Here, we report the same findings in tables. Table \ref{app_table_1} shows the performance in recovering unmoderated pairwise interactions (first row Figure \ref{fig_sim_results_part1}):

\begin{table}[H]
		\centering
	\scalebox{0.8} {
		\begin{tabular}{llllllllllllll}
			&             & \multicolumn{12}{c}{n}                                              \\ \hline
			&             & 30 & 43 & 63 & 92 & 133 & 193 & 280 & 407 & 591 & 858 & 1245 & 1808 \\ \hline
MNM (1)& SE & 0.01 & 0.00 & 0.01 & 0.02 & 0.06 & 0.14 & 0.28 & 0.59 & 0.83 & 0.97 & 1.00 & 1.00 \\ 
MNM (2)& SE  & 0.01 & 0.01 & 0.03 & 0.04 & 0.08 & 0.18 & 0.36 & 0.66 & 0.84 & 0.98 & 1.00 & 1.00 \\ 
MNM (3)& SE & 0.00 & 0.00 & 0.00 & 0.01 & 0.02 & 0.06 & 0.13 & 0.40 & 0.70 & 0.96 & 1.00 & 1.00 \\ 
MNM (1)& PR & 0.50 & 0.33 & 0.80 & 0.85 & 0.92 & 0.97 & 0.98 & 0.98 & 0.98 & 0.98 & 0.98 & 0.98 \\ 
MNM (2)& PR  & 0.50 & 0.57 & 0.82 & 0.90 & 0.90 & 0.97 & 0.98 & 0.97 & 0.97 & 0.95 & 0.95 & 0.94 \\ 
MNM (3)& PR &  &  &  & 0.80 & 0.89 & 0.97 & 0.98 & 1.00 & 1.00 & 1.00 & 0.99 & 1.00 \\ \hline
		\end{tabular}
	}
	\caption{Sensitivity and precision of all compared methods as a function of sample size $n$. The missing values for precision indicate that no edges were estimated in the 100 iterations of the simulation and hence precision could not be computed.}
	\label{app_table_1}
\end{table}

For the different methods, (1) indicates that the correct moderator was specified, (2) that moderators were specified in $p=13$ sequential models, and estimates were combined, and (3) indicates that all moderators are specified in a single model (see also Figures \ref{fig_sim_results_part1} and \ref{fig_sim_results_part2}).

Table \ref{app_table_2} shows the results on moderated pairwise interactions shown in the second row of Figure \ref{fig_sim_results_part1}:

\begin{table}[H]
		\centering
	\scalebox{0.8} {
		\begin{tabular}{llllllllllllll}
			&             & \multicolumn{12}{c}{n}                                              \\ \hline
			&             & 30 & 43 & 63 & 92 & 133 & 193 & 280 & 407 & 591 & 858 & 1245 & 1808 \\ \hline
MNM (1)& SE& 0.00 & 0.01 & 0.03 & 0.04 & 0.08 & 0.20 & 0.55 & 0.86 & 0.98 & 1.00 & 1.00 & 1.00 \\ 
MNM (2)& SE & 0.01 & 0.03 & 0.05 & 0.06 & 0.13 & 0.27 & 0.60 & 0.86 & 0.98 & 1.00 & 1.00 & 1.00 \\ 
MNM (3)& SE & 0.00 & 0.00 & 0.01 & 0.01 & 0.04 & 0.10 & 0.31 & 0.70 & 0.94 & 1.00 & 1.00 & 1.00 \\ 
MNM (1)& PR & 0.50 & 0.33 & 0.80 & 0.85 & 0.92 & 0.97 & 0.98 & 0.98 & 0.98 & 0.98 & 0.98 & 0.98 \\ 
MNM (2)& PR & 0.50 & 0.57 & 0.82 & 0.90 & 0.90 & 0.97 & 0.98 & 0.97 & 0.97 & 0.95 & 0.95 & 0.94 \\ 
MNM (3)& PR &  &  &  & 0.80 & 0.89 & 0.97 & 0.98 & 1.00 & 1.00 & 1.00 & 0.99 & 1.00 \\  \hline
		\end{tabular}
	}
	\caption{Sensitivity and precision of all compared methods as a function of sample size $n$. The missing values for precison indicate that no edges were estimated in the 100 iterations of the simulation and hence precision could not be computed.}
	\label{app_table_2}
\end{table}

Table \ref{app_table_3} shows the results on moderation effects with pairwise part shown in the first row of Figure \ref{fig_sim_results_part2}:

\begin{table}[H]
		\centering
	\scalebox{0.8} {
	\begin{tabular}{llllllllllllll}
		&             & \multicolumn{12}{c}{n}                                              \\ \hline
		&             & 30 & 43 & 63 & 92 & 133 & 193 & 280 & 407 & 591 & 858 & 1245 & 1808 \\ \hline
MNM (1)& SE & 0.00 & 0.00 & 0.00 & 0.02 & 0.04 & 0.14 & 0.36 & 0.82 & 0.99 & 1.00 & 1.00 & 1.00 \\ 
MNM (2)& SE & 0.00 & 0.00 & 0.00 & 0.02 & 0.05 & 0.20 & 0.46 & 0.79 & 0.95 & 0.98 & 0.98 & 0.98 \\ 
MNM (3)& SE & 0.00 & 0.00 & 0.00 & 0.01 & 0.01 & 0.10 & 0.25 & 0.71 & 0.96 & 1.00 & 1.00 & 1.00 \\ 
NCT (1)& SE & 0.00 & 0.00 & 0.00 & 0.01 & 0.03 & 0.10 & 0.22 & 0.40 & 0.67 & 0.90 & 1.00 & 1.00 \\ 
NCT (2)& SE & 0.00 & 0.00 & 0.01 & 0.01 & 0.02 & 0.10 & 0.22 & 0.41 & 0.65 & 0.88 & 0.99 & 1.00 \\ 
FGL (1)& SE & 0.00 & 0.00 & 0.02 & 0.06 & 0.18 & 0.33 & 0.71 & 0.92 & 0.99 & 1.00 & 1.00 & 1.00 \\ 
FGL (2& SE & 0.02 & 0.02 & 0.06 & 0.12 & 0.24 & 0.43 & 0.76 & 0.92 & 0.98 & 1.00 & 1.00 & 1.00 \\ 
MNM (1)& PR &  &  &  & 1.00 & 1.00 & 1.00 & 1.00 & 1.00 & 1.00 & 1.00 & 1.00 & 1.00 \\ 
MNM (2)& PR &  &  &  & 1.00 & 1.00 & 1.00 & 1.00 & 0.99 & 0.99 & 0.99 & 0.99 & 0.98 \\ 
MNM (3)& PR &  &  &  &  & 1.00 & 1.00 & 1.00 & 1.00 & 1.00 & 0.99 & 1.00 & 1.00 \\ 
NCT (1)& PR &  &  &  &  & 0.71 & 0.92 & 0.96 & 0.94 & 0.97 & 0.98 & 0.98 & 0.98 \\ 
NCT (2)& PR &  &  & 0.25 & 0.08 & 0.13 & 0.32 & 0.45 & 0.58 & 0.49 & 0.38 & 0.34 & 0.33 \\ 
FGL (1)& PR &  & 0.00 & 0.43 & 0.54 & 0.79 & 0.85 & 0.83 & 0.82 & 0.74 & 0.66 & 0.57 & 0.59 \\ 
FGL (2)& PR & 0.08 & 0.15 & 0.19 & 0.19 & 0.32 & 0.37 & 0.46 & 0.48 & 0.45 & 0.34 & 0.28 & 0.26 \\ \hline
	\end{tabular}
}
\caption{Sensitivity and precision of all compared methods as a function of sample size $n$. The missing values for precison indicate that no edges were estimated in the 100 iterations of the simulation and hence precision could not be computed.}
\label{app_table_3}
\end{table}

Table \ref{app_table_4} shows the results on moderation effects without pairwise part shown in the second row of Figure \ref{fig_sim_results_part2}:

\begin{table}[H]
	\centering
	\scalebox{0.8} {
		\begin{tabular}{llllllllllllll}
			&             & \multicolumn{12}{c}{n}                                              \\ \hline
			&             & 30 & 43 & 63 & 92 & 133 & 193 & 280 & 407 & 591 & 858 & 1245 & 1808 \\ \hline
MNM (1)& SE & 0.00 & 0.00 & 0.00 & 0.01 & 0.04 & 0.08 & 0.32 & 0.70 & 0.95 & 1.00 & 1.00 & 1.00 \\ 
MNM (2)& SE & 0.00 & 0.00 & 0.00 & 0.02 & 0.05 & 0.13 & 0.32 & 0.66 & 0.91 & 0.96 & 0.96 & 0.96 \\ 
MNM (3)& SE & 0.00 & 0.00 & 0.00 & 0.00 & 0.03 & 0.06 & 0.22 & 0.57 & 0.88 & 0.98 & 1.00 & 1.00 \\ 
NCT (1)& SE & 0.00 & 0.00 & 0.00 & 0.00 & 0.00 & 0.04 & 0.11 & 0.35 & 0.65 & 0.90 & 0.98 & 1.00 \\ 
NCT (2)& SE & 0.00 & 0.00 & 0.00 & 0.00 & 0.01 & 0.04 & 0.12 & 0.34 & 0.60 & 0.90 & 0.98 & 0.99 \\ 
FGL (1)& SE & 0.00 & 0.00 & 0.00 & 0.00 & 0.03 & 0.02 & 0.10 & 0.23 & 0.53 & 0.86 & 1.00 & 1.00 \\ 
FGL (2& SE & 0.00 & 0.01 & 0.00 & 0.00 & 0.03 & 0.02 & 0.10 & 0.24 & 0.54 & 0.86 & 1.00 & 1.00 \\ 
MNM (1)& PR &  &  &  & 1.00 & 1.00 & 1.00 & 1.00 & 1.00 & 1.00 & 1.00 & 1.00 & 1.00 \\ 
MNM (2)& PR &  &  &  & 1.00 & 1.00 & 1.00 & 1.00 & 0.99 & 0.99 & 0.99 & 0.99 & 0.98 \\ 
MNM (3)& PR &  &  &  &  & 1.00 & 1.00 & 1.00 & 1.00 & 1.00 & 0.99 & 1.00 & 1.00 \\ 
NCT (1)& PR  &  &  &  &  & 0.71 & 0.92 & 0.96 & 0.94 & 0.97 & 0.98 & 0.98 & 0.98 \\ 
NCT (2)& PR &  &  & 0.25 & 0.08 & 0.13 & 0.32 & 0.45 & 0.58 & 0.49 & 0.38 & 0.34 & 0.33 \\ 
FGL (1)& PR &  & 0.00 & 0.43 & 0.54 & 0.79 & 0.85 & 0.83 & 0.82 & 0.74 & 0.66 & 0.57 & 0.59 \\ 
FGL (2& PR & 0.08 & 0.15 & 0.19 & 0.19 & 0.32 & 0.37 & 0.46 & 0.48 & 0.45 & 0.34 & 0.28 & 0.26 \\ \hline
		\end{tabular}
	}
	\caption{Sensitivity and precision of all compared methods as a function of sample size $n$. The missing values for precision indicate that no edges were estimated in the 100 iterations of the simulation and hence precision could not be computed.}
	\label{app_table_4}
\end{table}

\end{document}